\documentclass{aa}
\usepackage{txfonts}
\usepackage{graphicx}
\usepackage{gensymb}   
\usepackage{epstopdf}
\usepackage{natbib}
\usepackage{amssymb,amsmath}

\begin{document}

\title{A LOFAR-IRAS cross-match study: the far-infrared radio correlation and the 150-MHz luminosity as a star-formation rate tracer}
\author{L. Wang\inst{1,2}, F. Gao\inst{1,2},  K. J. Duncan\inst{3}, W.L. Williams\inst{3},  M. Rowan-Robinson\inst{4}, J. Sabater\inst{5}, T. W. Shimwell\inst{3},  M. Bonato\inst{6, 7}, G. Calistro-Rivera\inst{3}, K. T. Chy\.zy\inst{8}, D. Farrah\inst{9, 10}, G. G{\"u}rkan\inst{11}, M.J.Hardcastle\inst{12}, I. McCheyne\inst{13},  I. Prandoni\inst{6}, S. C. Read\inst{12}, H.J.A. R{\"o}ttgering\inst{3},  D.J.B. Smith\inst{12}}
\institute{SRON Netherlands Institute for Space Research, Landleven 12, 9747 AD, Groningen, The Netherlands \email{l.wang@sron.nl} 
\and Kapteyn Astronomical Institute, University of Groningen, Postbus 800, 9700 AV Groningen, the Netherlands
\and Leiden Observatory, Leiden University, PO Box 9513, 2300 RA Leiden, the Netherlands
\and Astrophysics Group, Imperial College London, Blackett Laboratory, Prince Consort Road, London SW7 2AZ, UK
\and SUPA, Institute for Astronomy, Royal Observatory, Blackford Hill, Edinburgh, EH9 3HJ, UK
\and INAF - Istituto di Radioastronomia, and Italian ALMA Regional Centre, Via Gobetti 101, I-40129 Bologna, Italy
\and INAF - Osservatorio Astronomico di Padova, Vicolo Osservatorio 5, I-35122 Padova, Italy
\and Astronomical Observatory of the Jagiellonian University, ul. Orla 171, 30-244 Krak\'ow, Poland
\and Department of Physics and Astronomy, University of Hawaii, 2505 Correa Road, Honolulu, HI 96822, USA
\and Institute for Astronomy, 2680 Woodlawn Drive, University of Hawaii, Honolulu, HI 96822, USA
\and CSIRO Astronomy and Space Science, PO Box 1130, Bentley WA 6102, Perth, Australia
\and Centre for Astrophysics Research, School of Physics, Astronomy and Mathematics, University of Hertfordshire, College Lane, Hatfield AL10 9AB, UK
\and Astronomy Centre, Dept. of Physics \& Astronomy, University of Sussex, Brighton BN1 9QH, UK
}

\date{Received / Accepted}

\abstract
   {}
   {We aim to study the far-infrared radio correlation (FIRC) at 150 MHz  in the local Universe (at a median redshift $\left<z\right>\sim0.05$) and improve the use of the rest-frame 150-MHz luminosity, $L_{150}$, as a star-formation rate (SFR) tracer, which is unaffected by dust extinction.}
   {We cross-match the 60-$\mu$m selected Revised IRAS Faint Source Survey Redshift (RIFSCz) catalogue and the 150-MHz selected LOFAR value-added source catalogue in the Hobby-Eberly Telescope Dark Energy Experiment (HETDEX) Spring Field. We estimate $L_{150}$ for the cross-matched sources and compare it with the total infrared (IR) luminosity, $L_{\rm IR}$, and various SFR tracers.}
   {We find a tight linear correlation between $\log L_{150}$ and $\log L_{\rm IR}$ for star-forming galaxies, with a slope of 1.37. The median $q^{\rm IR}$ value (defined as the logarithm of the $L_{\rm IR}$ to $L_{150}$ ratio) and its rms scatter of our main sample are 2.14 and 0.34, respectively. We also find that $\log L_{150}$ correlates tightly with the logarithm of SFR derived from three different tracers, i.e., ${\rm SFR_{H\alpha}}$ based on the H$\alpha$ line luminosity, ${\rm SFR_{60}}$ based on the rest-frame 60-$\mu$m luminosity and ${\rm SFR_{IR}}$  based on $L_{\rm IR}$, with a scatter of 0.3 dex. Our best-fit relations between $L_{150}$ and these SFR tracers are, $\log L_{150} \ (L_\odot) = 1.35 (\pm 0.06) \times \log {\rm SFR_{H\alpha}} \ (M_{\odot}{\rm/yr)} + 3.20 (\pm 0.06)$, $\log L_{150} \  (L_\odot) = 1.31 (\pm 0.05) \times \log {\rm SFR}_{60} \ (M_{\odot}{\rm/yr)} + 3.14 (\pm 0.06)$, and  $\log L_{150} \ (L_\odot) = 1.37 (\pm0.05)\times \log {\rm SFR}_{\rm IR} \ (M_{\odot}{\rm/yr)} + 3.09 (\pm0.05)$, which show excellent agreement with each other.
}
   {}


\titlerunning{The far-infrared radio correlation}

\authorrunning{Wang et al.}

\maketitle

\section{Introduction}

The correlation between far-infrared (FIR) and radio luminosities in normal star-forming galaxies, i.e. without significant active galaxy nuclei (AGN) activity, was discovered by Helou et al. (1985) using data from the Infrared Astronomical Satellite (IRAS). It has been confirmed in many subsequent studies with facilities like the {\it Spitzer} Space Telescope, the Balloon-Borne Large Aperture Submillimeter Telescope (BLAST) and the {\it Herschel} Space Observatory (Condon 1992, Yun et al. 2001, Sargent et al. 2010, Bourne et al. 2011, Ivison et al. 2010a,b) and has continued to intrigue for its tightness  and extent over many orders of magnitude in luminosity.  This relationship between FIR and radio luminosity had been prefigured in earlier studies at 10 $\mu$m  by van der Kruit (1971, 1973), at 100 $\mu$m  by Rickard \& Harvey (1984), and at 60 $\mu$m using early-release IRAS data by Dickey \& Salpeter (1984) and de Jong et al. (1985).  Moreover, the FIR to radio correlation (FIRC) also seems to be more or less independent of redshift (e.g. Garrett 2002; Appleton et al. 2004; Ibar et al. 2008; Jarvis et al. 2010; Sargent et al. 2010; Bourne et al. 2011), although this is still an issue of intense debate as some studies do show evidence for redshift evolution (e.g. Seymour et al. 2009; Ivison et al. 2010a; Micha{\l}owski et al. 2010a, b; Magnelli et al. 2015; Basu et al. 2015; Delhaize et al. 2017). 

Harwit \& Pacini (1975) had proposed that the radio emission from star-forming galaxies could arise from supernova remnants (SNR) but Helou et al (1985) showed that SNR could account for less than 10$\%$ of the radio emission. Instead Helou et al (1985) suggested that relativistic electrons must leak out from SNR into the general magnetic field of the galaxy. This picture was later refined by Helou \& Bicay (1993). In an idealized calorimeter model first proposed by Voelk (1989), the cosmic ray electrons lose all of their energy before escaping the galaxy, which is optically thick to ultraviolet (UV)  photons. Assuming calorimetry, the logarithmic slope of the FIRC is equal to one (i.e. the FIRC is linear) as both the non-thermal synchrotron radiation and IR radiation (due to dust heated by UV photons) depend on the same star-formation rate (SFR). The calorimeter model, which may hold for starburst galaxies, was able to reproduce the tightness of the FIRC but also had several shortcomings. Alternative, more complex non-calorimetric models have also been proposed to explain the tight FIRC for normal star-forming galaxies (e.g. Bell 2003; Murgia et al. 2005; Thompson et al. 2006; Lacki, Thompson \& Quataert 2010; Schleicher \& Beck 2013). For example, the ``equipartition model'' by Niklas \& Beck (1997) was the first to predict that the logarithmic slope of the FIRC is different from one (i.e. the FIRC is non-linear) for normal star-forming galaxies. Although a detailed picture of the physical origin of the FIRC is still lacking, the basic understanding is that massive star formation is the driver of this correlation as UV photons from young stars heat dust grains which then radiate in the IR, and the same short-lived massive stars explode as supernovae which accelerate cosmic rays thereby contributing to non-thermal synchrotron emission in the radio.

An important application of the FIRC is the use of the radio continuum (RC) emission as a SFR tracer which (like the FIR-based SFR tracer) is not affected by dust extinction, as opposed to the often heavily obscured emission at  UV or optical wavelengths. Another advantage of using RC emission as a SFR tracer is that radio observations using interferometers from the ground can achieve much higher angular resolutions (arcsec or even sub-arcsec resolution) compared to single aperture IR telescopes in space. The {\it Herschel} space observatory was the largest IR telescope ever launched with a 3.5-metre primary mirror. The full width at half maximum (FWHM) of the {\it Herschel}-PACS beams are  (for the most common observing mode) 5.6$''$, 6.8$''$ and 10.7$''$ at 70, 100, and 160 $\mu$m, respectively\footnote{These values are taken from HERSCHEL-HSC-DOC-2151, version 1.0, February 28, 2017.} and the FWHM of the {\it Herschel}-SPIRE beams are 18.1$''$, 25.2$''$ and 36.6$''$ at 250, 350, and 500 $\mu$m, respectively (Swinyard et al. 2010).

The FIRC has been investigated mostly at GHz frequencies in the past, particularly at 1.4 GHz. For example, Yun et al. (2001) studied the NRAO Very Large Array (VLA) Sky Survey (NVSS) 1.4 GHz radio counterparts of IR galaxies selected from the IRAS Redshift survey out to $z\sim0.15$ and found the FIRC is well described by a linear relation over five orders of magnitude with a scatter of only 0.26 dex. Using 24 and 70 $\mu$m IR data from {\it Spitzer} and 1.4 GHz radio data from VLA, Appleton et al. (2004) found strong evidence for the universality of the FIRC out to $z\sim1$. Ivison et al. (2010b) studied the FIRC over the redshift range $0<z<2$ using multi-band IR data including observations from {\it Spitzer, Herschel} and SCUBA, and 1.4-GHz data from the VLA. They found no evidence for significant evolution of the FIRC with redshift. Using deep IR observations from {\it Herschel}  and deep 1.4-GHz VLA observations and Giant Metre-wave Radio Telescope (GMRT) 610-MHz observations in some of the most studied blank extragalactic fields, Magnelli et al. (2015) reported a moderate but statistically significant redshift evolution of the FIRC out to $z\sim2.3$. Thus, the overall conclusions are that there is a tight correlation between the FIR and radio luminosity at 1.4 GHz in the local Universe out to at least redshift $z\sim2$, but there is still ongoing debate over whether this correlation evolves with redshift.

With the advent of the LOw Frequency ARray (LOFAR; R{\"o}ttgering et al. 2011; van Haarlem et al. 2013) which combines a large field of view with high sensitivity on both small and large angular scales, we can now study the FIRC at lower frequencies where the contribution from thermal free-free emission is  even less important than at 1.4 GHz. Operating between 30 and 230 MHz, LOFAR offers complementary information to the wealth of data collected at higher frequencies. Using deep LOFAR 150-MHz observations in the 7 deg$^2$ Bo\"otes field (Williams et al. 2016), Calistro Rivera et al. (2017) studied the FIRC at 150 MHz from $z\sim0.05$ out to $z\sim2.5$. They found fairly mild redshift evolution in the logarithmic IR to radio luminosity ratio in the form of $q^{\rm IR} \sim (1+z)^{-0.22\pm0.05}$. However, if the FIRC is non-linear (i.e. the logarithmic slope is different from one), then it implies that the $q^{\rm IR}$ parameter would depend on luminosity.  Therefore the reported redshift dependence of $q^{\rm IR}$ may simply be a consequence of the non-linearity of the FIRC (Basu et al. 2015) as the mean SFR of galaxies is generally larger at higher redshifts (e.g., Hopkins \& Beacom 2006; Madau \& Dickinson 2014; Pearson et al. 2018; Liu et al. 2018; Wang et al. 2019). Based on LOFAR observations of the {\it Herschel} Astrophysical Terahertz Large Area Survey (H-ATLAS; Eales et al. 2010) 142 deg$^2$ North Galactic Pole (NGP) field (Hardscastle et al. 2016), G{\"u}rkan et al. (2018) found that a broken power-law (with a break around SFR $\sim1 M_{\odot}/\rm yr$) compared to a single power law is a better calibrator for the relationship between RC luminosity and SFR, possibly implying additional mechanisms for generating cosmic rays and/or magnetic fields. Also using LOFAR data in the NGP field, Read et al. (2018) found evidence for redshift evolution of the FIRC at 150 MHz. Heesen et al. (2019) studied the relation between radio emission and SFR surface density using spatially resolved LOFAR data of a few nearby spiral galaxies. They found a sublinear relation between the resolved RC emission and the SFR surface densities based on {\it GALEX} UV and $\it Spitzer$ 24 $\mu$m data.

The LOFAR Two-metre Sky Survey (LoTSS) is currently conducting a survey of the whole northern sky with a nominal central frequency of 150 MHz. The LoTSS First Data Release (DR1; Shimwell et al. 2019) contains a catalogue of over 325,000 sources detected over 425 deg$^2$ of the Hobby-Eberly Telescope Dark Energy Experiment (HETDEX) Spring Field, with a median sensitivity of 71 $\mu$Jy/beam and a resolution of $\sim6''$.  In this paper, we cross-match the LOFAR catalogue in the HETDEX Spring Field with the 60 $\mu$m selected Revised IRAS Faint Source Survey Redshift (RIFSCz; Wang et al. 2009, 2014a) Catalogue, which is constructed from the all-sky IRAS Faint Source Catalog (FSC), in order to study the FIRC in the local Universe and the use of the  rest-frame 150-MHz luminosity, $L_{150}$, as a SFR tracer.

There are several key differences between this study and the previous studies of Calistro Rivera et al. (2017), G{\"u}rkan et al. (2018) and Read et al. (2018) which were based solely on {\it Herschel} observations from either the {\it Herschel} Multi-tiered Extragalactic Survey (HerMES; Oliver et al. 2012) or H-ATLAS. First, the sky coverage of this study is at least three times larger than any previous studies, which means we can detect more rare sources such as ultra-luminous infrared galaxies (ULIRGs) with total IR luminosity ($L_{\rm IR}$) greater than $10^{12}L_{\odot}$ and SFR more than several hundred solar masses per year.  Secondly, the previous LOFAR studies relied on {\it Herschel} observations to determine $L_{\rm IR}$ of the LOFAR sources. The intrinsic 90\% completeness limit of the IRAS Faint Source Survey at 60 $\mu$m is $S_{60} = 0.36$ Jy (Wang \& Rowan-Robinson 2010). At the median redshift of our main sample $z\sim0.05$ (see Section 3.3), this flux limit corresponds to a 60-$\mu$m luminosity of $L_{60}\sim10^{10.27}L_{\odot}$, or equivalently $L_{\rm IR}\sim10^{10.5}L_{\odot}$, based on the median ratio of $L_{60}$ to $L_{\rm IR}$ using the IR spectral energy distribution (SED) templates from Chary \& Elbaz (2001). In comparison, the H-ATLAS $5\sigma$ limit, including both confusion and instrumental noise, is 37 mJy (Valiante et al. 2016) at  250 $\mu$m which is the most sensitive band. At $z\sim0.05$, this flux limit corresponds to a 250-$\mu$m luminosity of $L_{250}\sim10^{8.84}L_{\odot}$, or equivalently $L_{\rm IR}\sim10^{10.2}L_{\odot}$, based on the median ratio of $L_{250}$ to $L_{\rm IR}$ using the Chary \& Elbaz (2001) templates. Therefore, the IRAS observations are only a factor of $\sim2$ shallower than the H-ATLAS survey. Finally,  the IRAS photometric bands sample the peak of the dust SED for the  IR luminous galaxies in the local Universe. In comparison, the {\it Herschel}-SPIRE bands sample the Rayleigh-Jeans regime of the SED. Due to the lack of photometric bands covering the peak of the IR SED, both G{\"u}rkan et al. (2018) and Read et al. (2018) focused on the relation between the $L_{250}$ and $L_{150}$, rather than between $L_{\rm IR}$ and $L_{150}$. Most of the sources in the RIFSCz lie at redshift below 0.1 and thus provide an excellent local benchmark. The median redshift of our main sample is $z\sim0.05$. In comparison, the lowest redshift bin in the Calistro Rivera et al. (2017) study has a median redshift of 0.16. The sample used in G{\"u}rkan et al. (2018) and Read et al. (2018) covers the redshift range at $z<0.25$, with a median redshift of 0.1.

The paper is structured as follows. In Section 2, we introduce the two main datasets (and their associated multi-wavelength data) in our analysis, namely the RIFSCz catalogue and the LOFAR value-added catalogue (VAC) in the HETDEX Spring Field. The construction of the LOFAR-RIFSCz cross-matched sample and its basic properties such as its wavelength coverage and redshift distribution are summarised in Section 3. In Section 4, we present the main results of our study, the FIRC at both 1.4 GHz and 150 MHz and the correlation between the rest-frame 150-MHz luminosity and various SFR tracers. Finally, we give our conclusions in Section 5. Throughout the paper, we assume a flat $\Lambda$CDM universe with $\Omega_m=0.3$,  $\Omega_{\Lambda}=0.7$,  and $H_0=70$ km s$^{-1}$ Mpc$^{-1}$. We adopt a Kroupa (2001) initial mass function (IMF) unless stated otherwise.

\section{Data}

\subsection{The RIFSCz catalogue}

The Revised IRAS Faint Source Survey Redshift (RIFSCz) Catalogue (Wang et al. 2009, 2014a; Rowan-Robinson \& Wang 2015) is composed of galaxies selected from the IRAS Faint Source Catalog (FSC) over the whole sky at  Galactic latitude |b| > $20^{\circ}$. RIFSCz incorporates data from {\it GALEX}, the Sloan Digital Sky Survey (SDSS; York et al. 2000), the Two Micron All Sky Survey (2MASS; Skrutskie et al. 2006), the Wide-field Infrared Survey Explorer (WISE; Wright et al. 2010), and Planck all-sky surveys (Planck Collaboration I 2013) to give wavelength coverage from 0.36-1380 $\mu$m.  At a 60-$\mu$m flux density of $S_{60} > 0.36$ Jy, which is the 90\% completeness limit of the FSC, 93\% of RIFSCz sources have optical or NIR counterparts with spectroscopic or photometric redshifts (Wang et al. 2014a). Spectroscopic redshifts are compiled from the SDSS spectroscopic DR10 survey (Ahn et al. 2014), the 2MASS Redshift Survey (2MRS; Huchra et al. 2012), the NASA/IPAC Extragalactic Database (NED), the PSC Redshift Survey (PSCz; Saunders et al. 2000), the 6dF Galaxy Survey, and the FSS redshift survey (FSSz; Oliver, PhD thesis). Photometric redshifts are derived by applying the template-fitting method  used to construct the SWIRE Photometric Redshift Catalogue (Rowan-Robinson et al. 2008 and references therein). Six galaxy templates and three QSO templates are used. For sources with at least 8 photometric bands and with reduced $\chi^2<3$, the percentage of catastrophic outliers, i.e. $(1 + z_{\rm phot})$ differs from $(1 + z_{\rm spec})$ by more than 15\%, is 0.17\%  and the rms accuracy is 3.5\% after exclusion of these outliers. IR SED templates are fitted to the mid- and far-IR data, following the methodology of Rowan-Robinson et al (2005, 2008) and as in Wang \& Rowan-Robinson (2009), with a combination of two cirrus templates, three starburst templates and an AGN dust torus template. The total IR luminosity $L_{\rm IR}$ (integrated between 8 and 1000 $\mu$m) is estimated based on the fitted templates.

The methodology of Rowan-Robinson et al (2008) is followed to calculate stellar masses and SFR. Briefly, the rest-frame 3.6-$\mu$m luminosity is estimated and converted to stellar mass using the mass-to-light ratio derived from stellar synthesis models. To estimate SFR,  the conversion recipes of Rowan-Robinson et al. (1997) and Rowan-Robinson (2001) are used
\begin{equation}
{\rm SFR}_{60} \ (M_{\odot} {\rm /yr}) = 2.2 \eta^{-1} 10^{-10} L_{60} \ (L_{\odot})
\end{equation}
where $\eta$ is the fraction of UV light absorbed by dust, taken as 2/3. The SFRs are calculated for a Salpeter IMF (1955) between 0.1 and 100 $M_{\odot}$. To convert to Kroupa (2001) IMF, we divide the values by 1.5.  We can also estimate SFR based on the total IR luminosity $L_{\rm IR}$ following the widely used recipe of Kennicutt (1998) after converting to Kroupa IMF,
\begin{equation}
{\rm SFR}_{\rm IR} \ (M_{\odot} {\rm /yr})   = 10^{-10} L_{\rm IR} \ (L_{\odot}).
\end{equation}
In principle, the formula of Eq. (2) is only suitable for dusty starburst galaxies in which all of the radiation from young stars is assumed to be absorbed by dust and subsequently re-emitted in the IR. In practice, Eq. (2) has been found to also apply to normal galaxies (e.g. Rosa-Gonz{\'a}lez, Terlevich \& Terlevich 2002; Charlot et al. 2002). The explanation is that there are two competing effects, which are overestimation in SFR caused by assuming all of the IR luminosity arises from recent star formation (as opposed to old stellar populations) and underestimation in SFR caused by neglecting the possibility that some of the young stellar radiation is not absorbed by dust. It is a coincidence that these two effects cancel out (e.g. Inoue 2002; Hirashita, Buat \& Inoue 2003).

For sources in the RIFSCz which have been cross-matched to SDSS DR 10, we  also have SFR estimates based on the H$\alpha$ line luminosity, ${\rm SFR_{H\alpha}}$, corrected for dust attenuation and aperture effects provided in the MPA-JPU database (Brinchmann et al. 2004).

\subsection{The LOFAR survey}

Exploiting the unique capabilities of LOFAR (van Haarlem et al. 2013), LoTSS is an ongoing sensitive, high-resolution, low-frequency (120-168 MHz) radio survey of the northern sky and is described in Shimwell et al. (2017). LoTSS provides the astrometric precision needed for accurate and robust identification of optical and NIR counterparts (e.g. McAlpine et al. 2012) and a sensitivity that, for typical radio sources, is superior to previous wide area surveys at higher frequencies such as the NRAO VLA Sky Survey (NVSS; Condon et al. 1998) and Faint Images of the Radio Sky at Twenty-Centimeters (FIRST; Becker, White, \& Helfand 1995) and is similar to forthcoming higher frequency surveys such as the Evolutionary Map of the Universe (EMU; Norris et al. 2011), and the APERture Tile In Focus survey (e.g. Rottgering et al. 2011). The primary observational objectives of LoTSS are to reach a sensitivity of less than 100$\mu$Jy/beam at an angular resolution, defined as the FWHM of the synthesised beam, of $\sim6''$ across the whole northern hemisphere.
 
The LoTSS First Data Release (DR1) presents 424 deg$^2$ of RC observations over the HETDEX Spring Field (10$^h$45$^m$00$^s$ $<$ right ascension$<$ 15$^h$30$^m$00$^s$ and 45$^{\circ}$00$^{'}$00$^{''}$ $<$ declination $<$ 57$^{\circ}$ 00$^{'}$ 00$^{''}$) with a median sensitivity of 71$\mu$Jy/beam and a resolution of $6''$, resulting in a catalogue with over 325,000 sources. Shimwell et al. (2019) estimated that the positional accuracy of the catalogued sources is better than $0.2''$. The VAC includes optical cross matches and photometric redshifts for the LOFAR sources. The procedure of cross-matching to currently available optical and mid-IR photometric surveys is presented in Williams et al. (2019). Photometric redshifts (phot-$z$) are estimated using a combination of template fitting methods and empirical training based methods (Duncan et al. 2019). The overall scatter and outlier fraction in the phot-$z$ is 3.9\% and 7.9\%, respectively. Following Read et al. (2018), we calculate the K-corrected 150-MHz luminosity assuming a spectral shape of $S_{\nu} \propto \nu^{-\alpha}$, where the spectral index $\alpha=0.71$ (Condon 1992; Mauch et al. 2013).

\section{The RIFSCz-LOFAR cross-matched sample}

In order to cross-match the IRAS sources in the RIFSCz catalogue and LOFAR sources in the HETDEX Spring Field, we take a combined approach of the closest match method and the likelihood ratio (LR) method as detailed below. 

\subsection{The closest match method}

\begin{figure}
\centering
\includegraphics[height=2.6in,width=3.65in]{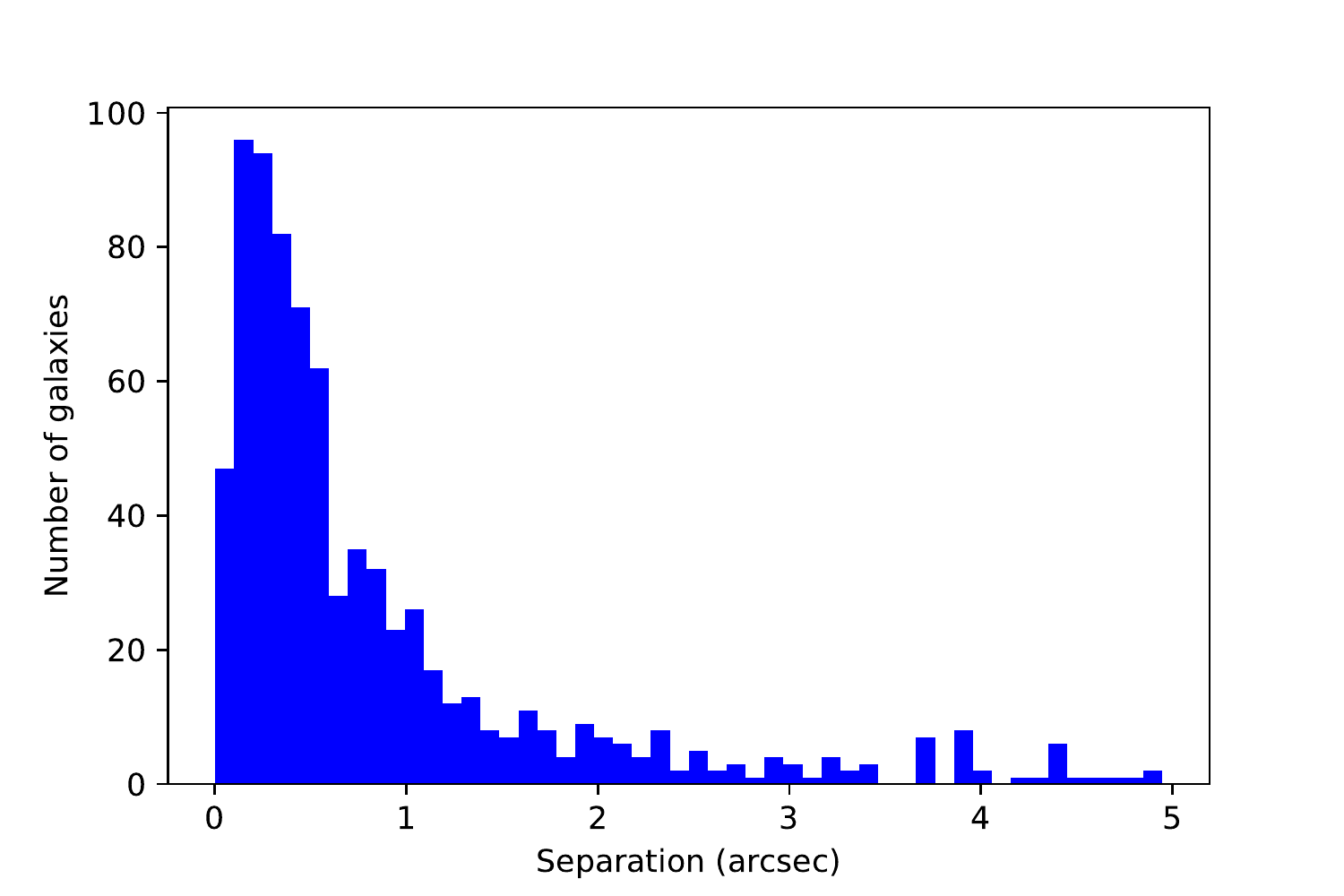}
\includegraphics[height=2.6in,width=3.65in]{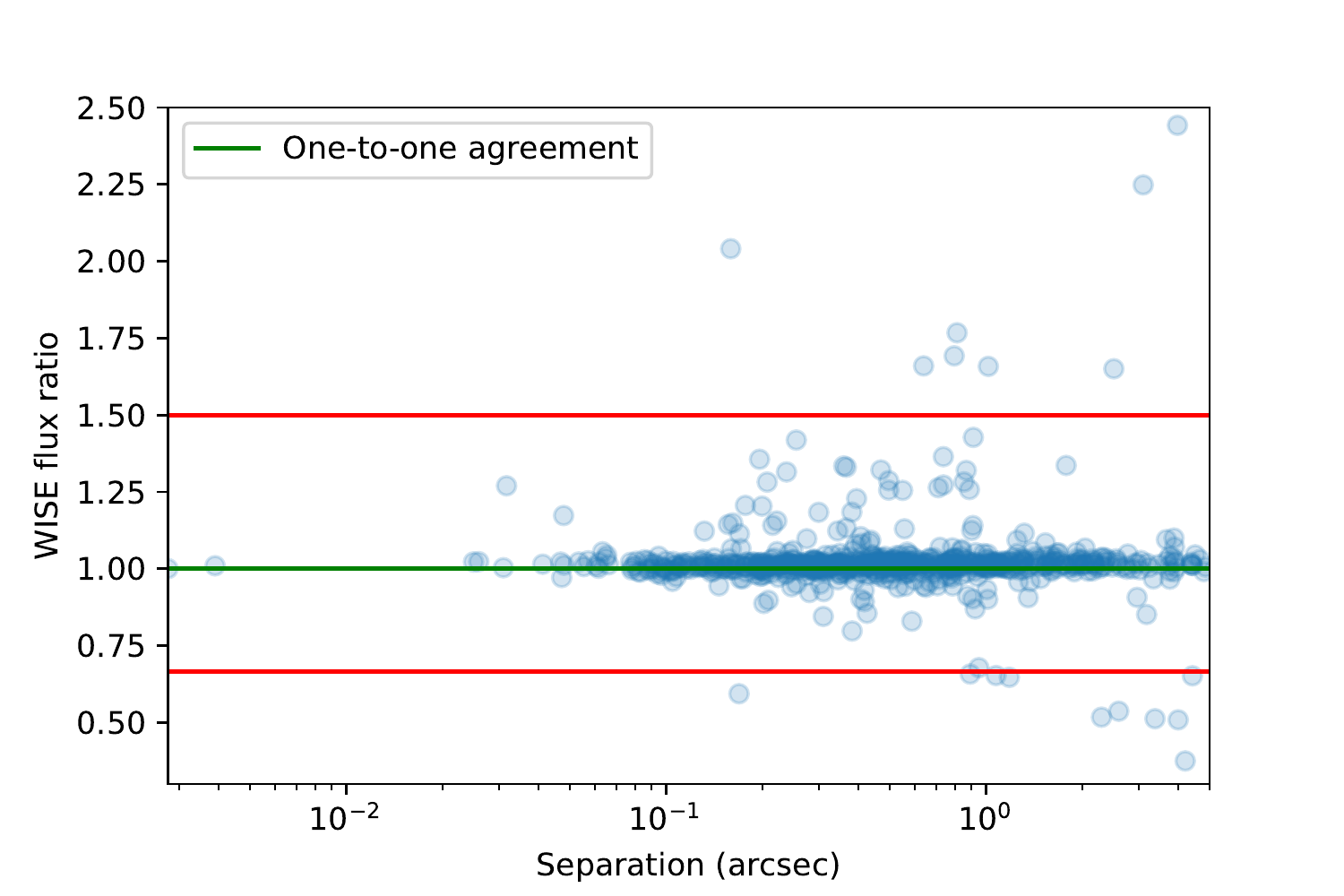}
\includegraphics[height=2.6in,width=3.65in]{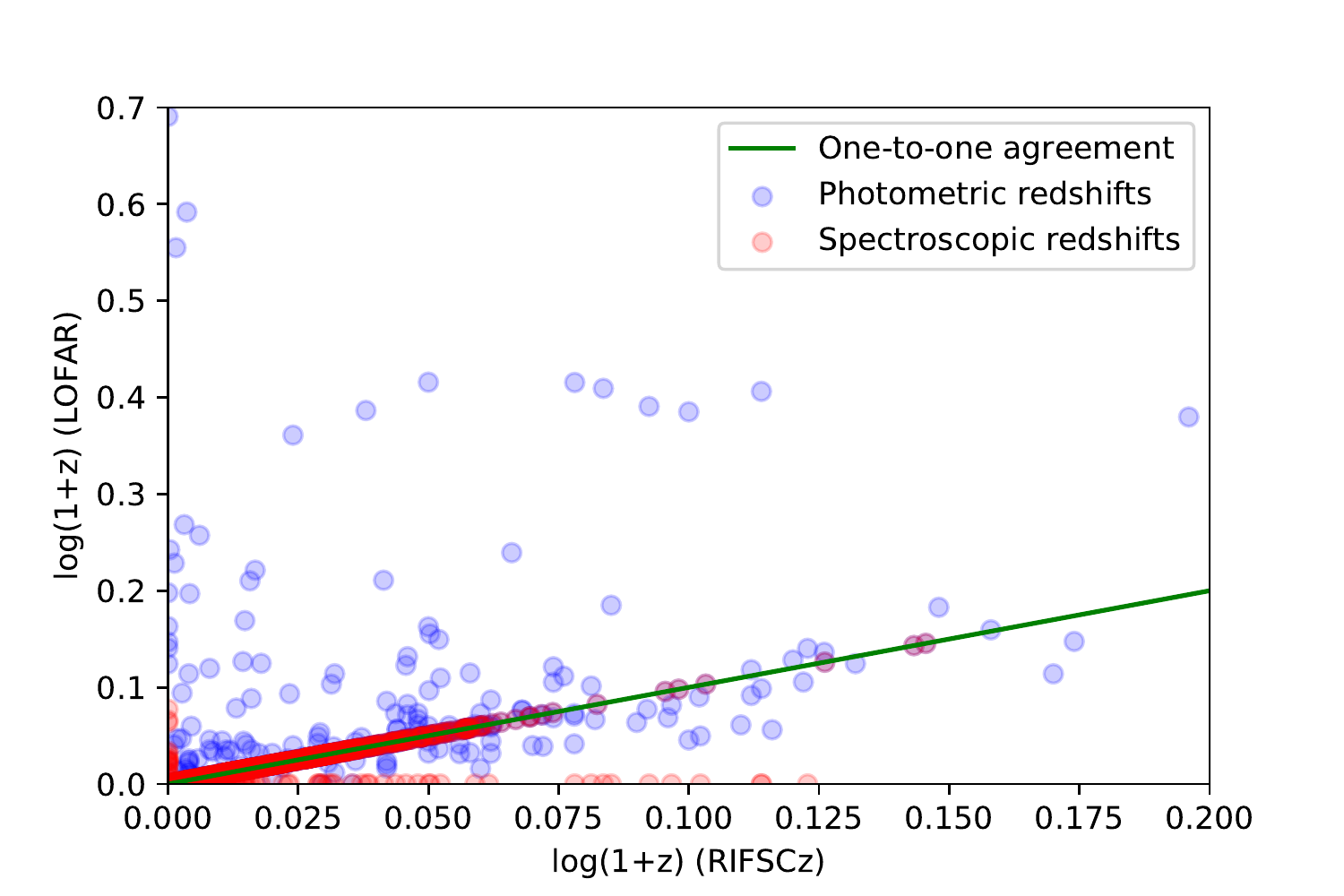}
\caption{Top: The distribution of positional separations of sources matched between RIFSCz and LOFAR. Middle: Comparison of WISE W1 flux for sources listed in RIFSCz and in LOFAR.  Sources inside the two horizontal red lines have good WISE  flux agreement (i.e., the difference is within a factor of 1.5). Bottom: Comparison of redshifts compiled in the RIFSCz and LOFAR VAC.}
\label{comp1}
\end{figure}

For IRAS  sources in the RIFSCz which are matched to sources detected at other wavelengths (e.g., the SDSS optical bands or the WISE IR bands), we choose the closest LOFAR match within a $5''$ searching radius which results in a cross-matched sample of 771 sources\footnote{The positions given in the RIFSCz catalogue correspond to the positions of the multi-wavelength cross-id matched to the IRAS sources, prioritised in the order of SDSS, 2MASS, WISE, NED and IRAS FSC.}. The conservative choice of $5''$ for the searching radius is mainly motivated by the FWHM of the LOFAR beam, although we note that the positional uncertainty is much smaller than that (Shimwell et al. 2019). Only one source has two possible matches (one located at $1.8''$ away and the other at $4.4''$ away). The top panel of Fig.~\ref{comp1} shows that the majority of the matches have positional differences well within $1''$, consistent with what we expect from the positional accuracies of LOFAR, SDSS and WISE (York et al. 2000; Wright et al. 2010; Shimwell et al. 2019).

The middle panel of Fig.~\ref{comp1} compares the WISE W1 fluxes at 3.4 $\mu$m provided by the cross-id in both the RIFSCz and LOFAR catalogues. The excellent agreement for the vast majority of sources demonstrates that we have the same id  for most of the RIFSCz-LOFAR matched sources.  Some sources have fairly different WISE fluxes which indicate potential problems with the cross-ids (between RIFSCz and LOFAR, between RIFSCz and WISE, or between LOFAR and WISE). Therefore, we exclude a total of 22 sources for which the WISE flux ratio from the two catalogues differs by more than a factor of 1.5. 

The bottom panel of Fig.~\ref{comp1} compares redshifts  provided for the RIFSCz-LOFAR matched sources from both catalogues, after excluding the 22 sources that could be erroneous matches. The spectroscopic redshifts (spec-$z$) show excellent agreement. 15 sources that have no spec-$z$ in the RIFSCz now have a spec-$z$ from LOFAR (based on the SDSS DR14). 71 sources that have no spec-$z$ from LOFAR but have a spec-$z$ from RIFSCz\footnote{In the RIFSCz, the recommended spec-z and flags are 1=SDSS DR10, 2=PSCz, 3=FSSz, 4=6dF, 5=NED and 6=2MRS, prioritised as NED$>$SDSS$>$2MRS$>$PSCz$>$FSSz$>$6dF. These spectroscopic surveys (except SDSS) are not used in the construction of the LOFAR VAC.}. The origin for these new spec-$z$ are NED (54 out of 71), SDSS (2 out of 71), PSCz (3 out of 71), FSSz (12 out of 71).  A generally good agreement can be found between the phot-$z$ estimates from both catalogues. In some cases, the LOFAR phot-$z$ tend to be higher than the phot-$z$ from the RIFSCz. We have studied 39 cases where the phot-$z$ estimates differ by more than 0.2 and found that the higher LOFAR phot-$z$ are likely to be erroneous because they would imply unrealistically high optical luminosity. Therefore, we adopt a priority order of redshift estimates as follows: spec-$z$ from RIFSCz (652 sources), followed by spec-$z$ from the LOFAR VAC (15 sources), followed by phot-$z$ from RIFSCz (76 sources), and finally phot-$z$ from LOFAR (6 sources).

To summarise, we select the sources with good WISE flux agreement (749 out of 771) and call this our ``main sample''. All of the sources in the main sample have redshift estimates.  Out of 749 sources, 581 sources (78\%) have spec-$z$ from both RIFSCz and the LOFAR VAC. As discussed in the paragraph above, the two spec-$z$ values are in perfect agreement with each other. We refer to this subset of the main sample as the ``main spec-$z$ sample'' which is our most robust sample with no ambiguity in the multi-wavelength cross-id. If we include the 15 new spec-$z$ from LOFAR and the new 71 spec-$z$ from RIFSCz, then we increase the sample size to 667 galaxies (89\%) and we refer to this subset as the ``main joint spec-$z$ sample''. Finally, 82 sources (11\%) have phot-$z$. We refer to this subset of the main sample as the ``main phot-$z$ sample''.

\subsection{The likelihood ratio method (LR)}

\begin{figure}
\centering
\includegraphics[height=2.6in,width=3.65in]{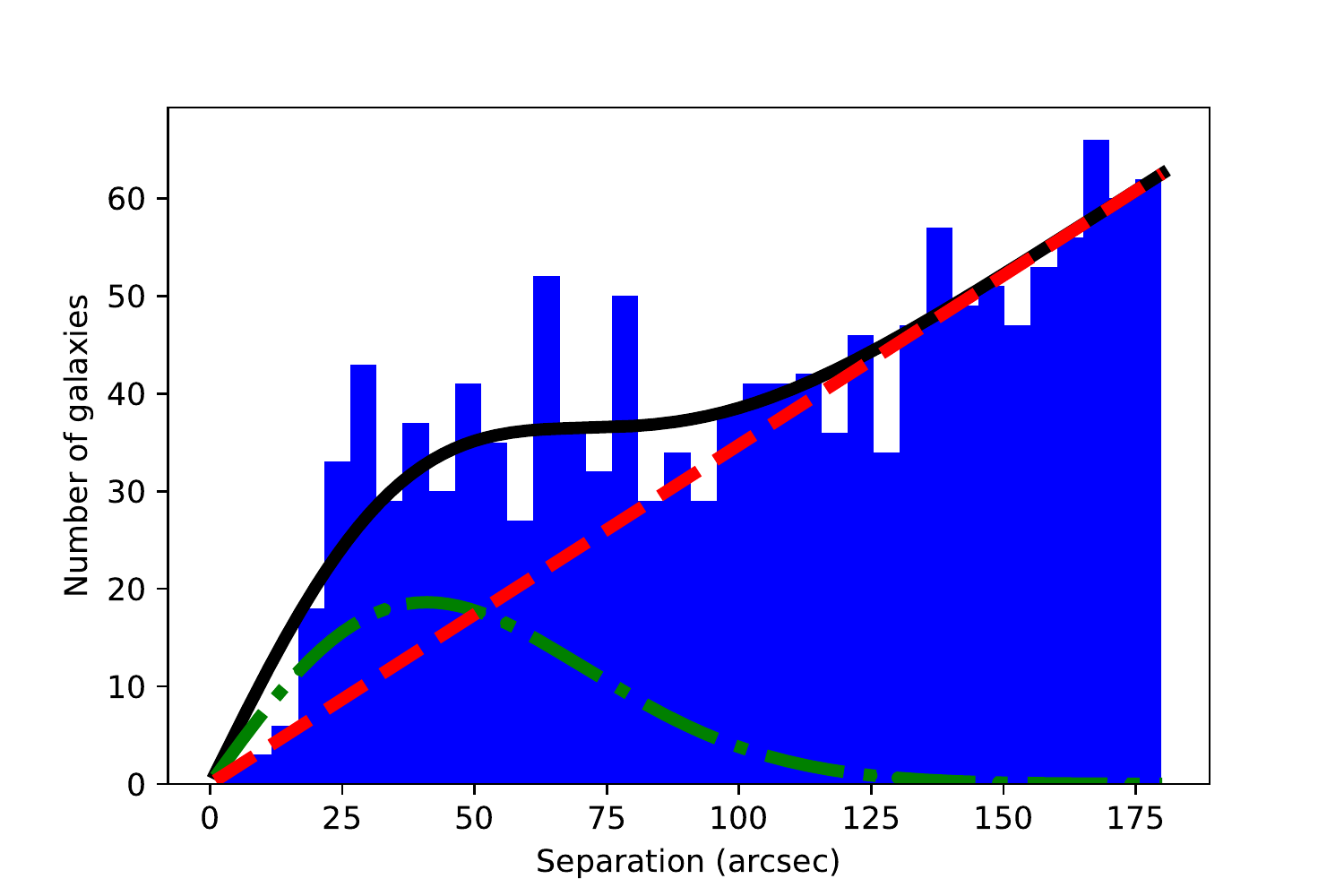}
\caption{The distribution of radial offsets between the RIFSCz sources (which only have IRAS observations) and LOFAR sources by selecting all matches within 3$'$. The radial distribution of the random associations is plotted as the red dashed line, while the radial distribution of the true counterparts is shown as the green dot dashed line. The black solid line is the sum of the two.}
\label{onlyiras}
\end{figure}

\begin{figure}
\centering
\includegraphics[height=2.6in,width=3.65in]{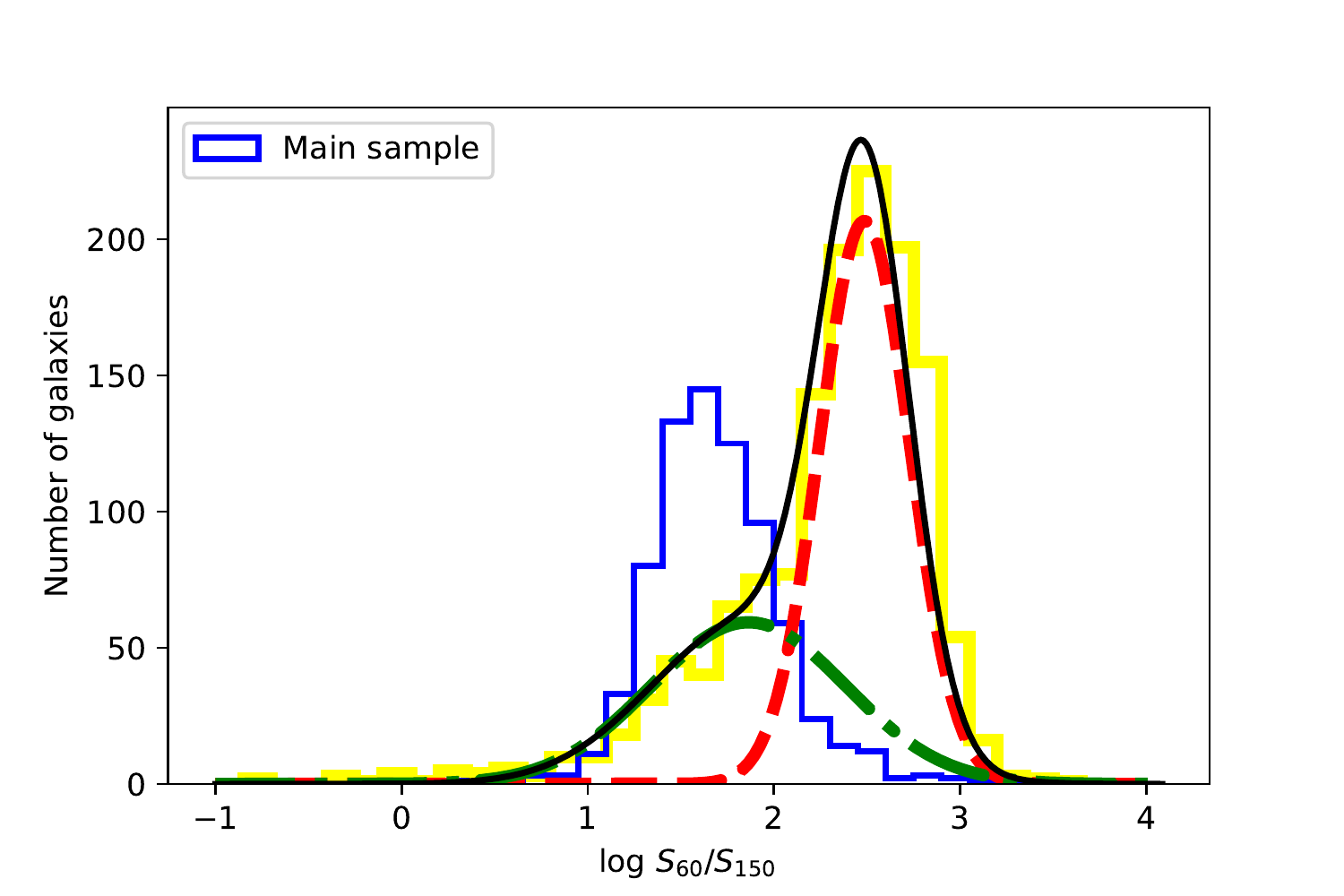}
\caption{The 60-$\mu$m $-$ 150-MHz colour distribution of all matches within 3$'$ between the RIFSCz sources (which only have IRAS observations) and LOFAR sources (yellow histogram). The dot-dashed Gaussian represents the inferred colour distribution of the true counterparts and the dashed Gaussian represents that of the random associations. The black solid line is the sum of the two. The colour distribution of the main sample is shown as the blue histogram.}
\label{onlyiras_colour}
\end{figure}

For IRAS sources in the RIFSCz which have not been matched to sources at other wavelengths and therefore only have IRAS positions\footnote{These IRAS only sources can be selected by applying FLAG position = 5 in the RIFSCz catalogue. Around 19\% of the sources in the RIFSCz catalogue have only IRAS observations.}, we adopt an LR method (Sutherland \& Saunders 1992; Brusa et al. 2007; Wang \& Rowan-Robinson 2010; Chapin et al. 2011; Wang et al. 2014a) in order to match them with LOFAR sources. The accurate LOFAR positions would then allow these IRAS only sources to be matched with optical or NIR sources. The LR technique compares the probability of a true counterpart with the probability of a chance association, as a function of 60 $\mu$m to 150 MHz flux ratio $S_{60}/S_{150}$ and radial offset $r$. Assuming the probability of true counterpart and random association is separable in $\log_{10}(S_{60}/S_{150})$ (or $C_{60-150}$ as a shorthand) and $r$, we can write
\begin{equation}
LR = \frac{Prob_{true}(C_{60-150}, r)}{Prob_{random}(C_{60-150}, r)} = \frac{q(C_{60-150}) E f(r)dC dr}{p(C_{60-150}) \rho b(r)dC dr},
\end{equation}
where $q(C_{60-150})$ and $p(C_{60-150})$ are the colour distributions of the true counterparts and random matches respectively, and $f(r)$ and $b(r)$ are the positional distributions of the true counterparts and random associations respectively.

To derive the positional distribution of the true counterparts $f(r)$, we assume a symmetric Gaussian distribution as a function of orthogonal positional coordinates. Therefore, $f(r)$ can be written as a Rayleigh radial distribution,
\begin{equation}
f(r)dr = \frac{r}{\sigma_r^2} \exp(-r^2/2\sigma_r^2)dr,
\end{equation}
where the scale parameter,  $\sigma_r$, is where $f(r)$ peaks and $\int_0^{\infty} f(r)dr = 1$. The positional distribution of random associations can be written as,
\begin{equation}
b(r)dr = 2\pi rdr,
\end{equation}
assuming a constant surface density of background LOFAR sources uncorrelated with IRAS sources. 

In Fig.~\ref{onlyiras}, we plot the distribution of radial offsets between the IRAS-only RIFSCz sources and LOFAR sources by selecting all matches within $3'$, which contains both the true counterparts and the random associations. We fit our model
\begin{equation}
N(r) dr = E \times f(r) dr + \rho \times b(r) dr,
\end{equation}
to the observed histogram to determine the best-fit parameters to be $E = 251.24\pm34.28$, $\sigma_r = 40.92''\pm3.59''$ and $\rho = 0.0111\pm0.0005$. This is consistent with what we expect based on the positional accuracy of IRAS sources. The angular resolution of IRAS varied between about $0.5'$ at 12 $\mu$m to about $2'$ at 100 $\mu$m. The positional accuracy of the IRAS sources depends on their size, brightness and SED but is usually better than $20''$ (1-$\sigma$). A histogram of the angular separations between IRAS positions and the NED positions can be found in Wang \& Rowan-Robinson (2009).  

In  Fig.~\ref{onlyiras_colour}, we plot the 60-$\mu$m $-$ 150-MHz colour distribution of all matches within $3'$ between the RIFSCz sources (which only have IRAS observations) and LOFAR sources. These matches contain both true and random associations. We assume that this colour distribution can be fit by two Gaussian distributions. We also plot the colour distribution of the RIFSCz-LOFAR matches from the main sample  discussed in Section 3.1. It is clear that there are systematic differences in median values and widths between the green dot-dashed line and the blue histogram. This is caused by the difference in the redshift ranges (see discussions in Section 3.3).

Having derived the positional and colour probability distributions of the true and random associations, we can now calculate the LR for every possible match based on its positional separation and IR-to-radio colour. So, for every RIFSCz object with more than one LOFAR counterpart within $3'$, we select the match with the highest LR\footnote{A total of 9 IRAS sources only have one LOFAR match within 3$'$. For these sources, we simply select the only LOFAR match.}. We also impose a minimal LR threshold to ensure the false identification rate is no more than 10\%. The LR threshold is derived as follows:
\begin{enumerate}
\item[$-$] First, we calculate the LR distribution of matches between a randomised RIFSCz and a randomised LOFAR VAC. The randomised catalogues are generated by randomly re-arranging the flux measurements of the sources, while keeping the positions unchanged.

\item[$-$]  Then, we compare the LR distribution of the matches between the randomised catalogues with that of the matches between the original catalogues (i.e. before randomisation). 

\item[$-$]  Finally, we set the minimal LR threshold to that above which the number of random matches is 10\% of the number of matches between the original catalogues.

\end{enumerate}

In total, 141 galaxies are matched between RIFSCz and LOFAR using the LR method. Out of the 141 galaxies, 112 galaxies have multi-wavelength optical and NIR data in the LOFAR VAC which are then used in the phot-$z$ estimation procedure discussed in Section 2.1. We refer to this subset of 112 galaxies matched between RIFSCz and LOFAR using the LR method as the ``second sample''. 79 galaxies in the second sample have spec-$z$ from the VAC. We refer this as the second spec-$z$ sample and the rest of the galaxies as the second phot-$z$ sample.

\subsection{Summary of the cross-matched sample}

\begin{table}
\centering
\caption{The number of sources in the main sample of the cross-matched RIFSCz-LOFAR sample (749 sources in total) by wavelength coverage. For the IRAS fluxes, we require moderate- or high-quality flux measurement. The exception is the IRAS 60 $\mu$m band where all sources have high-quality flux measurement.}\label{table:selection}
\begin{tabular}{lll}
\hline
Wavelength ($\mu$m) & Survey & Number of sources \\
\hline
3.4&WISE& 748\\
4.6&WISE& 748\\
12 &WISE& 748 \\
12    &     IRAS & 59\\
22 &WISE& 748\\
25    &     IRAS & 135\\
60    &     IRAS & 749\\
65 & AKARI & 194\\
90 & AKARI & 205\\
100    &     IRAS & 452\\
140 & AKARI & 196\\
160 & AKARI & 171 \\
350 & PLANCK &56\\
550& PLANCK &54\\
850& PLANCK &50\\
1380 & PLANCK &36\\
\hline
\end{tabular}
\end{table}

\begin{table}
\centering
\caption{The number of sources in the second sample of the cross-matched RIFSCz-LOFAR sample (112 sources in total) by wavelength coverage.}\label{table:selection}
\begin{tabular}{lll}
\hline
Wavelength ($\mu$m) & Survey & Number of sources \\
\hline
3.4&WISE& 107\\
4.6&WISE& 106\\
12 &WISE& 94 \\
12    &     IRAS & 1\\
22 &WISE& 83\\
25    &     IRAS & 12\\
60    &     IRAS & 112\\
65 & AKARI & 2\\
90 & AKARI & 2\\
100    &     IRAS & 51\\
140 & AKARI & 1\\
160 & AKARI & 2 \\
350 & PLANCK &3\\
550& PLANCK &3\\
850& PLANCK &2\\
1380 & PLANCK &2\\
\hline
\end{tabular}
\end{table}

\begin{figure}
\centering
\includegraphics[height=2.6in,width=3.65in]{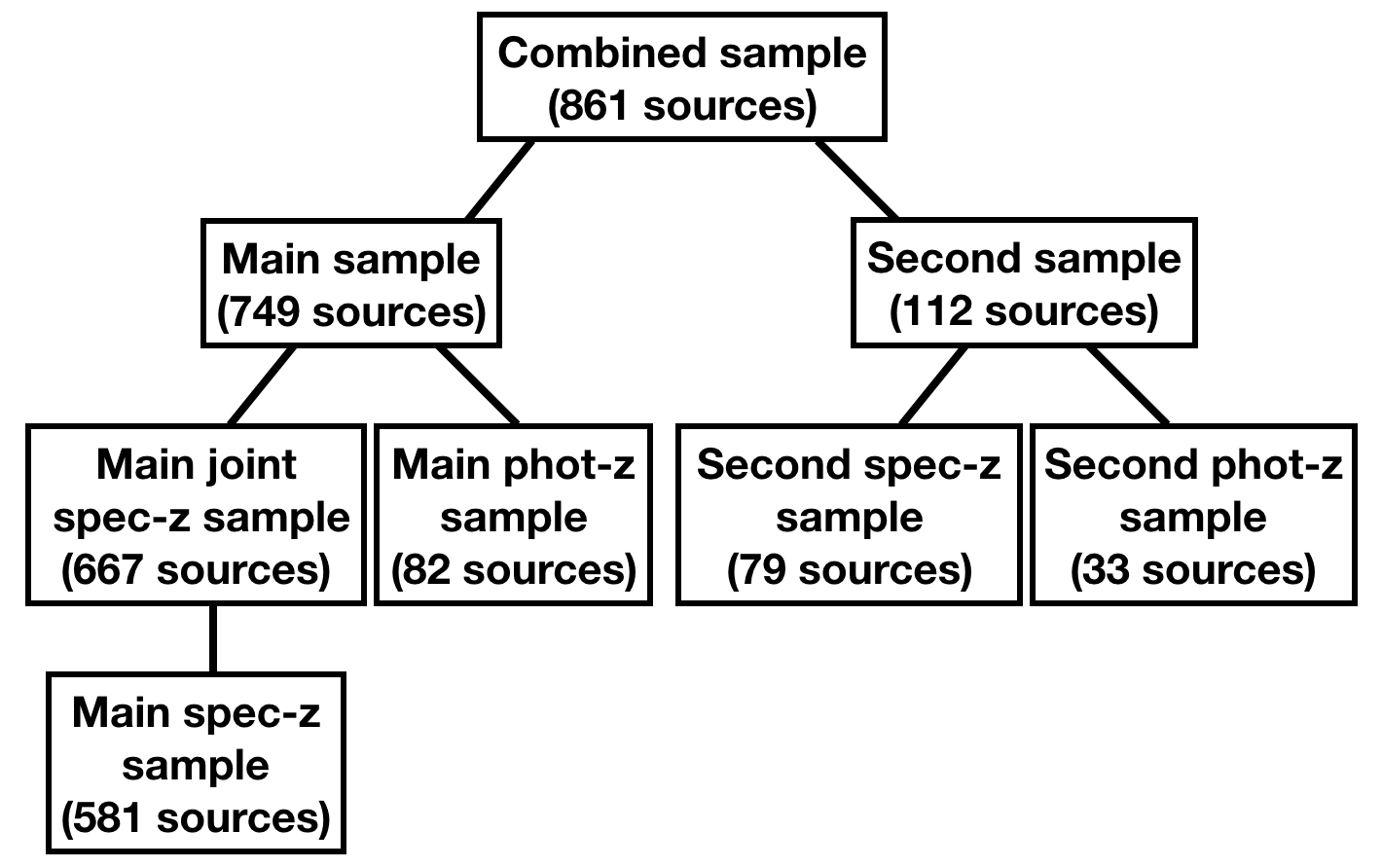}
\caption{A schematic view of our RIFSCz-LOFAR matched sample.}
\label{sum}
\end{figure}

\begin{figure}
\centering
\includegraphics[height=2.6in,width=3.65in]{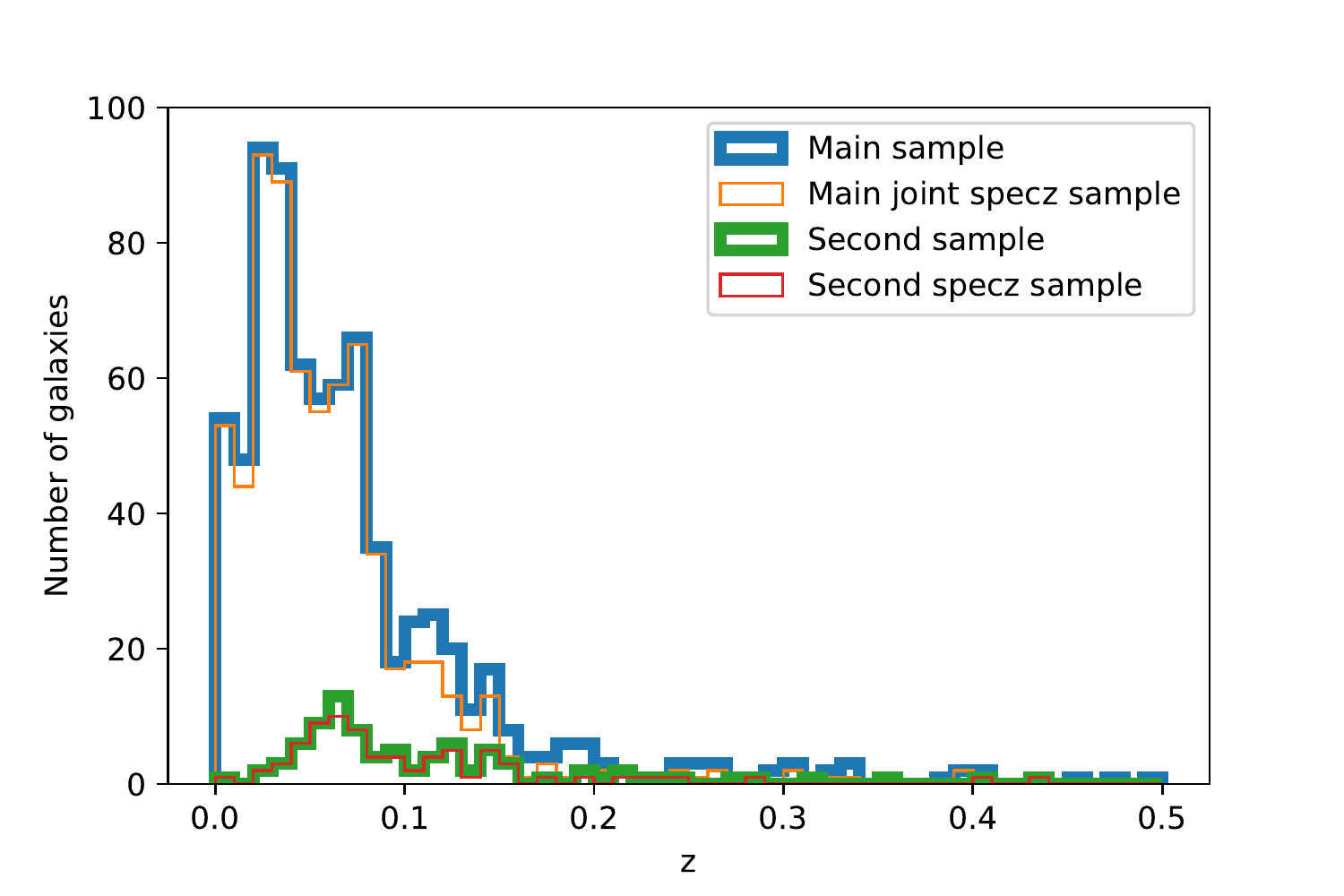}
\includegraphics[height=2.6in,width=3.65in]{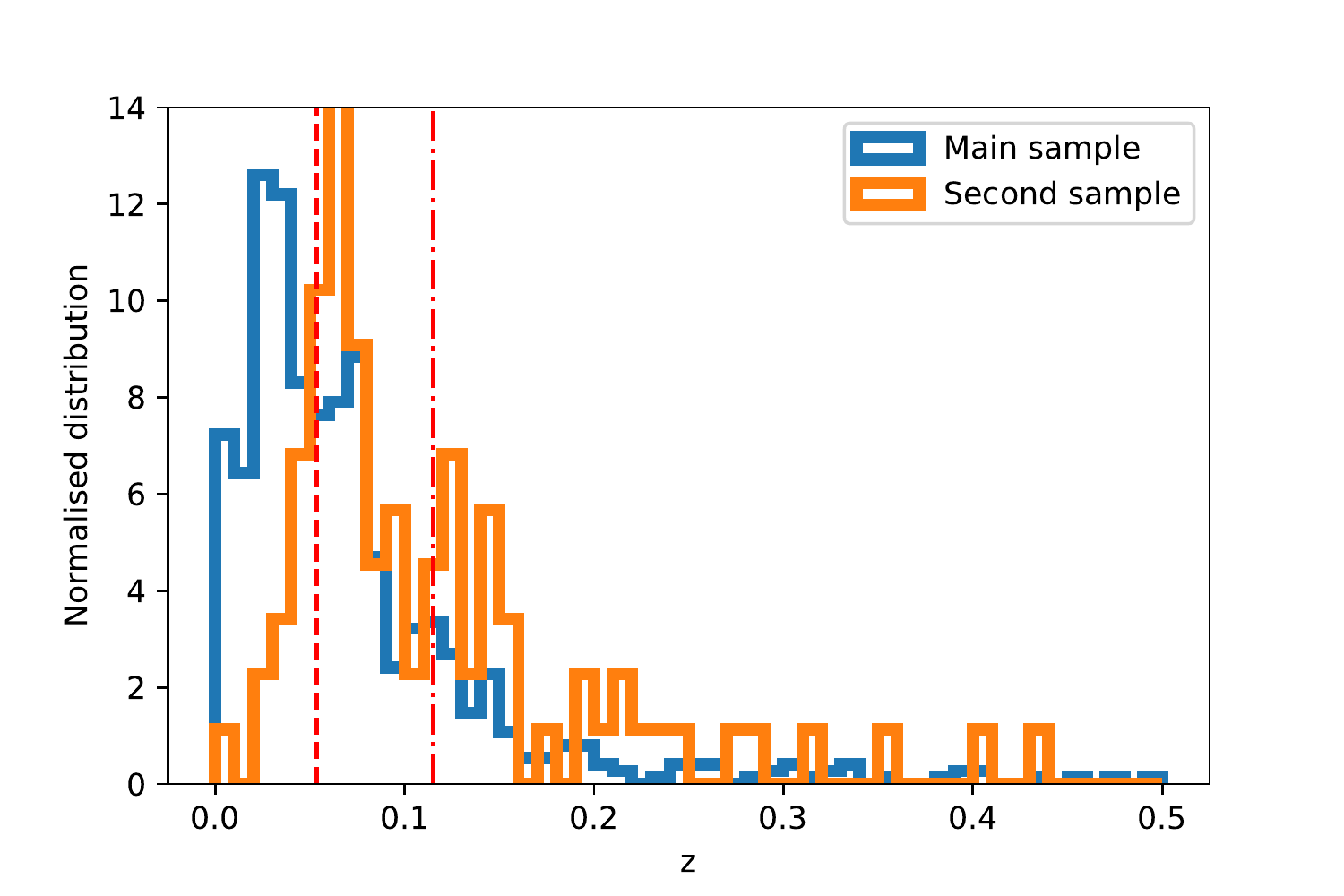}
\caption{Top: The redshift distribution of the RIFSCz-LOFAR cross-matched sample. Bottom: The normalised distributions (i.e. the integral of the distribution is 1). The median redshifts of the main sample and the second sample are 0.05 (indicated by the dashed line) and 0.12 (the dot-dashed line), respectively.}
\label{Nz}
\end{figure}

Fig.~\ref{sum} shows a schematic view of our RIFSCz-LOFAR matched sample. The combined sample of 861 sources is a combination of the main sample (generated using the closest match method) and the second sample (generated using the likelihood ratio method). Both samples are divided into subsamples depending on whether the sources have spec-$z$ or phot-$z$. In the main sample, there are a total of 581 sources with spec-$z$ from both RIFSCz and LOFAR which we refer to as the main spec-$z$ sample. An additional 86 sources have spec-$z$ from either LOFAR or RIFSCz which form the main joint spec-$z$ sample after combining with the main spec-$z$ sample. The top panel in Fig.~\ref{Nz} shows the redshift distribution of the cross-matched RIFSCz-LOFAR sample. Most galaxies have spec-$z$. The majority of our sources lie at $z<0.1$. The bottom panel shows the normalised distribution  to bring out the contrast in the redshift distribution. The median redshift of the main sample and the second sample is 0.05 and 0.12, respectively.

Table 1 shows the number of sources in the main sample by IR wavelength coverage (i.e. the number of sources detected at a given IR wavelength).  Most sources have been matched to WISE. For the IRAS fluxes, the flux quality is classified as high (NQ $=3$), moderate (NQ $=2$) or upper limit (NQ $=1$). We require flux quality flag NQ $>1$ to avoid upper limits. The exception is the 60-$\mu$m band. All sources in the RIFSCz have high-quality flux measurement in the 60-$\mu$m band. A small fraction also have AKARI flux measurement out to 160 $\mu$m. A very small number of sources also have Planck measurements at 250, 550, 850 and 1380 $\mu$m. Table 2 shows the number of sources in the second sample by IR wavelength coverage. Again, most sources have been matched to WISE. As the second sample is generally at higher redshift than the main sample, the IR SED coverage is poorer especially at the longer wavelengths from AKARI and Planck.

\section{Results}

Given that the FIRC has been very well studied at 1.4 GHz (see Section 1),  in this section we first study the FIRC at 1.4 GHz and compare with previous studies. Then we focus on the FIRC at 150 MHz and possible variations with respect to redshift.  After that, we investigate the use of the 150-MHz luminosity density as a SFR tracer.

\subsection{The FIR-radio correlation at 1.4 GHz}

\begin{figure}
\centering
\includegraphics[height=2.6in,width=3.65in]{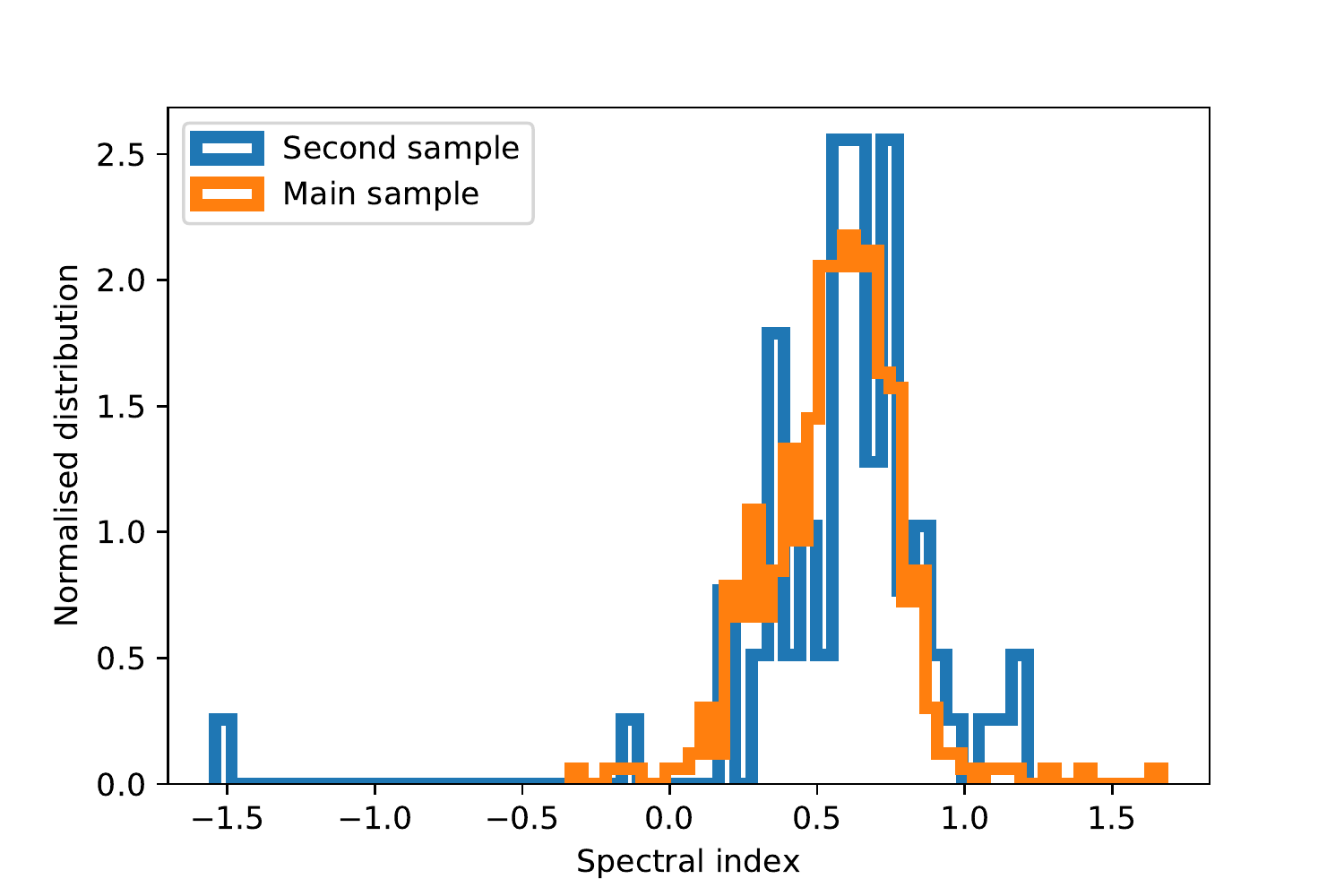}
\caption{The normalised distribution of the radio spectral index between 150 MHz and 1.4 GHz.}
\label{spectral_index}
\end{figure}


\begin{figure}
\centering
\includegraphics[height=2.6in,width=3.65in]{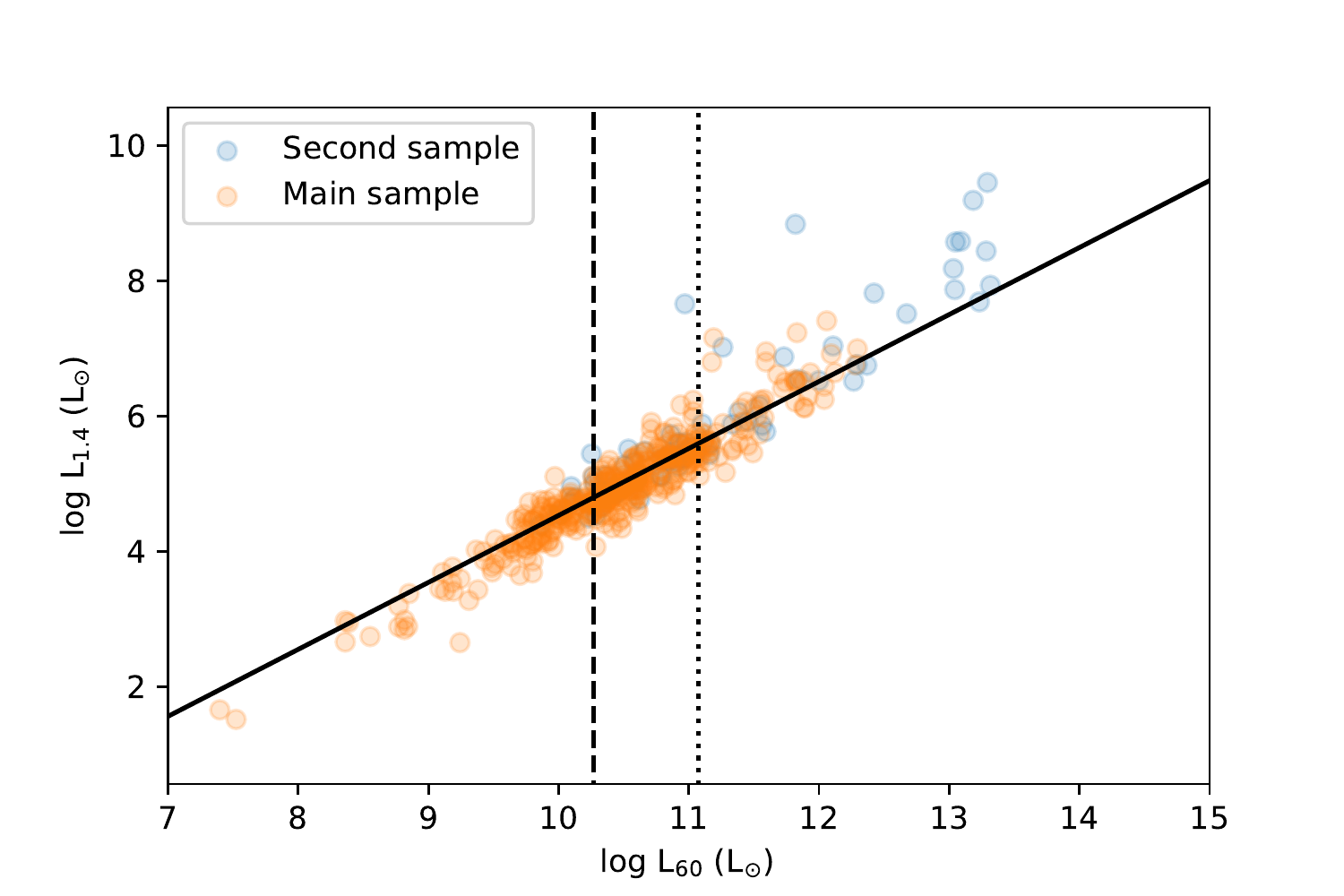}
\includegraphics[height=2.6in,width=3.65in]{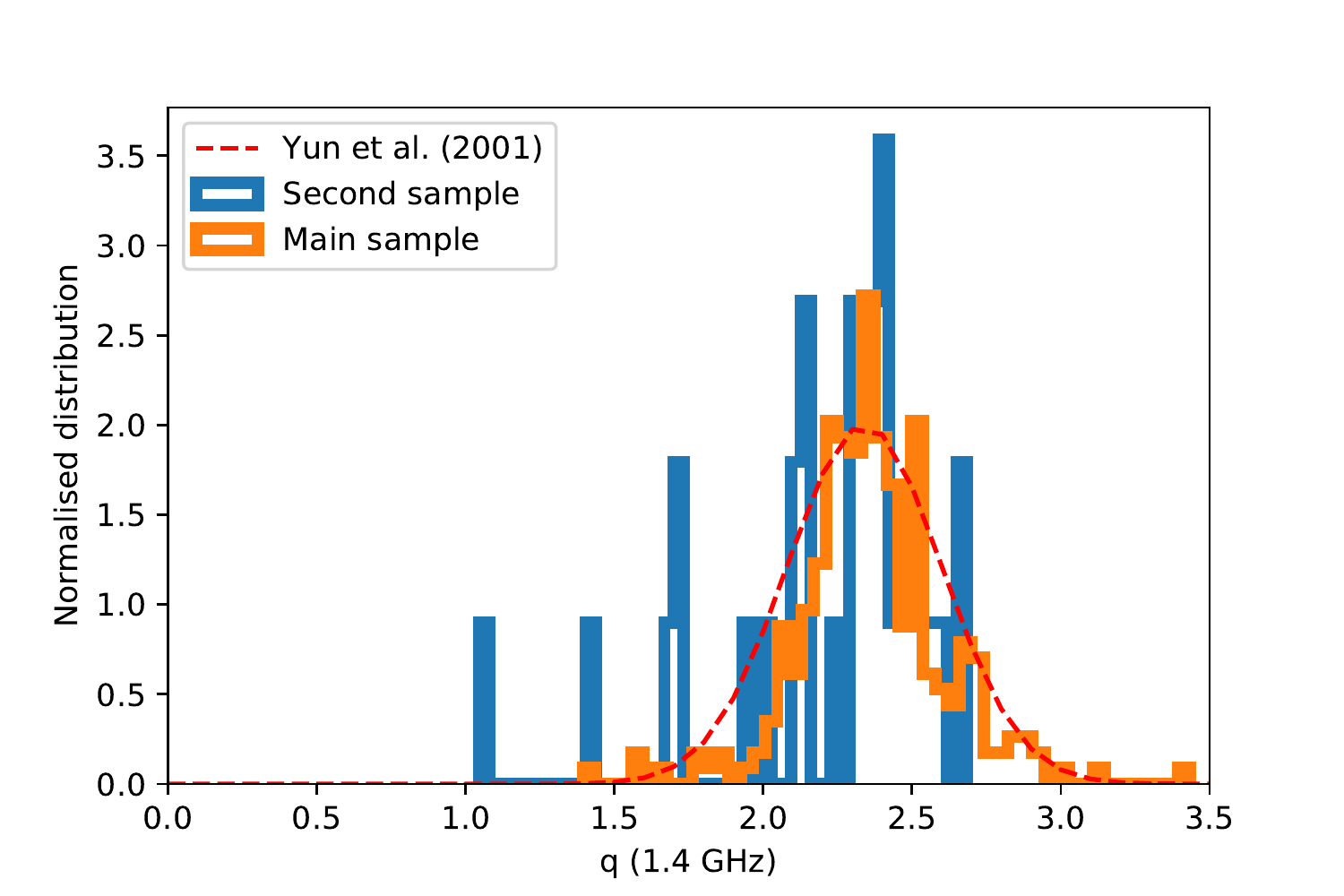}
\caption{Top:  The 1.4 GHz radio luminosity plotted against the IRAS 60 $\mu$m luminosity. The vertical dashed line indicates the 90\% completeness limit at the median redshift of the main sample. The vertical dotted line indicates the 90\% completeness limit of the second sample. The solid line is the Yun et al. (2001)  relation. Bottom: Histogram of $q~(\rm 1.4 GHz)$ values, derived using Eq. (8),  using only sources with NQ $>1$ at 100 $\mu$m. The dashed line is a Gaussian distribution with mean and standard deviation set to 2.34 and 0.26 respectively, which are values found by Yun et al. (2001).}
\label{FIRC1400}
\end{figure}

{We obtained the 1.4GHz FIRST survey catalogue (14Dec17 version) which contains 946,432 sources observed from the 1993 through 2011 observations\footnote{\url{http://sundog.stsci.edu/first/catalogs.html}}. The FIRST detection limit is 1 mJy over most of the survey area. The angular resolution of FIRST is $\sim5''$, similar to LOFAR. We cross-matched FIRST with LOFAR by selecting the closest match within $3''$. 412 matches were found with the main sample and 79 matches were found with the second sample.  We derive the radio spectral index by following
\begin{equation}
\alpha_{\nu_1}^{\nu_2} = \frac{\log(S_{\nu_1}/S_{\nu_2})}{\log(\nu_2/\nu_1)}
\end{equation}
where $\nu_1$ = 150 MHz and $\nu_2$ = 1400 MHz. Figure ~\ref{spectral_index} shows the histogram of the derived spectral index values. We do not find a significant difference between the main sample and the second sample. The median value of the spectral index and scatter for the main sample are 0.58 and 0.22, respectively. The median value and scatter for the second sample are 0.64 and 0.35, respectively. These values are very similar to the spectral index found in Sabater et al. (2019) using the galaxies overlapping between the SDSS DR7 and LoTSS. Sabater et al. (2019) also showed that their spectral index value (median value 0.63) is probably biased to lower values for low luminosity galaxies due to selection biases in the shallower 1.4 GHz sample compared to the low-frequency LOFAR data (which misses sources with steeper radio spectra). The spectral index values found in our samples are also likely to be biased to lower values compared to the canonical value of 0.71 (see Section 2.2) because of the shallower 1.4 GHz data.

In the top panel of Fig.~\ref{FIRC1400}, we plot the 1.4-GHz radio luminosity against the IRAS 60-$\mu$m luminosity.  The vertical dashed line indicates the 90\% completeness limit $L_{60} \sim 10^{10.27} L_{\odot}$ at the median redshift $z\sim0.05$ of the main sample.  The vertical dotted line indicates the 90\% completeness limit $L_{60} \sim 10^{11.08} L_{\odot}$ at the median redshift $z\sim0.12$ of the second sample. In comparison, the detection limit of FIRST of around 1 mJy corresponds to a 1.4 GHz luminosity $L_{1.4}\sim10^{4.34} L_{\odot}$ at $z\sim0.05$ and $L_{1.4}\sim10^{5.14} L_{\odot}$ at $z\sim0.12$. Yun et al. (2001) studied a sample of IRAS sources with $S_{60} > 2$ Jy and found that over 98\% of their sample follow a linear FIRC over five orders of magnitude in luminosity with a scatter of only 0.26 dex. We overplot their best-fit relation (with a slope of 0.99) in the top panel in Fig.~\ref{FIRC1400}. Most of our sources seem to follow the Yun et al. (2001) relation. Some sources in our second sample show deviations from the Yun et al. (2001) relation. However, the second sample is much smaller and less reliable.

Because the FIRC has a slope of unity, it can also be examined with  the ``$q$'' parameter, which is the logarithmic FIR to radio flux ratio and is commonly defined as (e.g., Helou et al. 1985; Condon  et al. 1991; Yun et al. 2001),
\begin{equation}
q~(\rm 1.4 GHz) = \log\left(\frac{S_{\rm FIR}}{3.75\times10^{12}}\right) - \log(S_{\rm 1.4})
\end{equation}
where $S_{1.4}$ is the observed 1.4 GHz flux density in units of W m$^{-2}$ Hz$^{-1}$ and 
\begin{equation}
S_{\rm FIR} = 1.26 \times 10^{-14} (2.586\times S_{60} + S_{100}) \ {\rm W m^{-2}}
\end{equation}
where $S_{60}$ and $S_{100}$ are the IRAS 60 and 100 $\mu$m flux densities in Jy (Helou et al. 1988). In the bottom panel Fig.~\ref{FIRC1400}, we plot the $q~(\rm 1.4 GHz)$ values derived for our sample, using only sources for which NQ $>1$ at 100 $\mu$m. This requirement on moderate- or high-quality flux measurement at 100 $\mu$m reduces the sizes of the main and second sample to 452 and 51, respectively (see Table 1 and Table 2). We do not see a significant difference between the main sample and the second sample. The median $q~(\rm 1.4 GHz)$ value and rms scatter for the main sample are 2.35 and 0.25 respectively, while the median $q~(\rm 1.4 GHz)$ value and scatter for the second sample are 2.34 and 0.35 respectively, using sources for which NQ $>1$ at 100 $\mu$m. This indicates that there is no significant redshift evolution in the $q~(\rm 1.4 GHz)$ value although the redshift range probed by our sample is probably too small to detect this.  We over-plot a Gaussian distribution with mean and standard deviation set to the values in Yun et al. (2001). The distributions of $q~(\rm 1.4 GHz)$ of our samples agree well with the Yun et al. (2001) distribution.

\begin{figure}
\centering
\includegraphics[height=2.6in,width=3.65in]{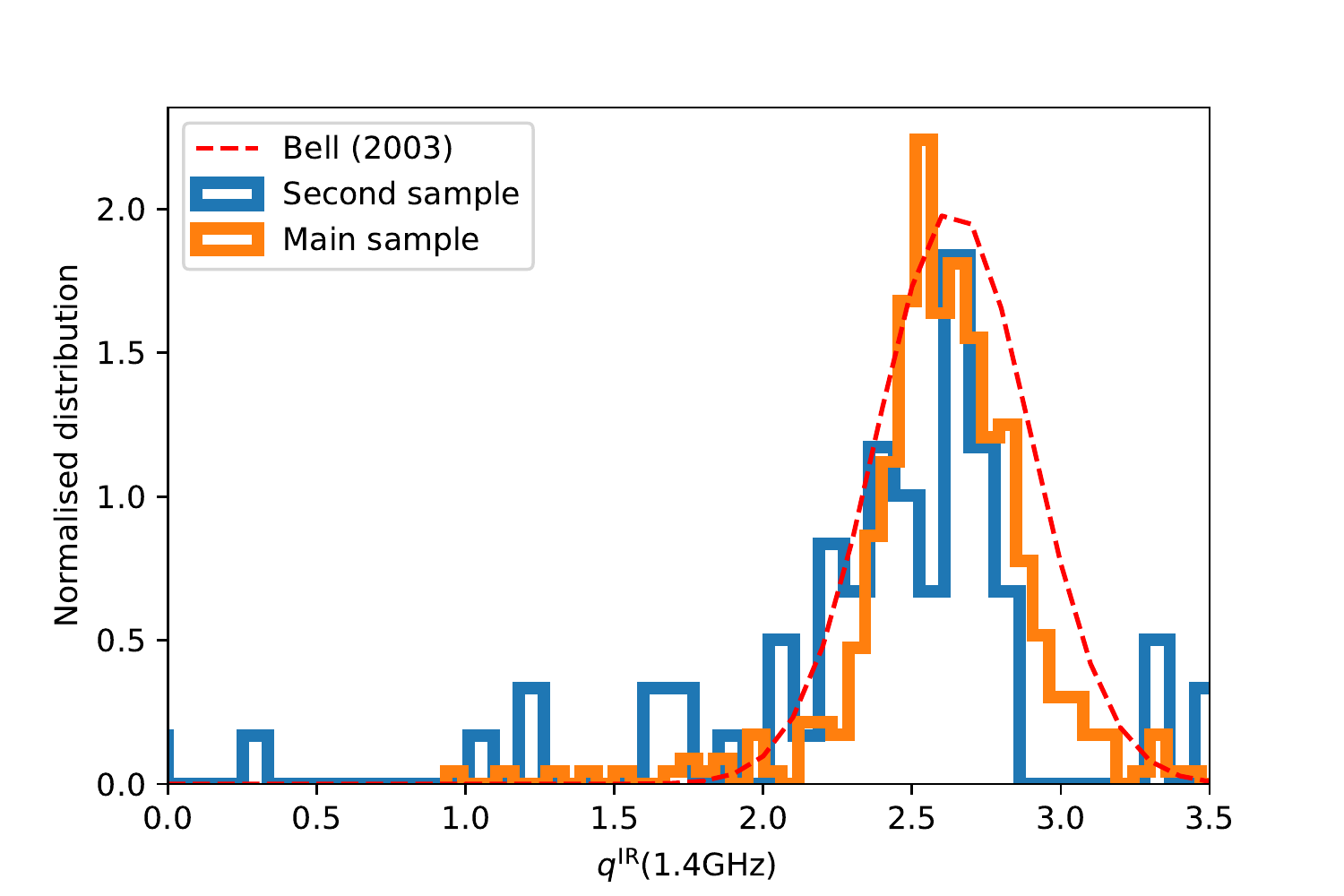}
\caption{Histogram of $q^{\rm IR}$ values at 1.4 GHz, derived using Eq. (10). The dashed line is a Gaussian with its mean and standard deviation set to 2.64 and 0.26 respectively, which are values found by Bell (2003).}
\label{FIRC1400_bell}
\end{figure}

Bell (2003) proposed an alternative definition of $q$  using the total IR to radio luminosity ratio, 
\begin{equation}
q^{\rm IR}~(\rm 1.4 GHz) = \log\left( \frac{L_{\rm IR}/(3.75\times10^{12}Hz)}{L_{\rm 1.4}}\right)
\end{equation}
where  $L_{\rm 1.4}$ is the 1.4-GHz luminosity. In Fig.~\ref{FIRC1400_bell}, we plot the distribution of the $q^{\rm IR}~(\rm 1.4 GHz)$  values of our sample. Bell (2003) found a median value of 2.64 and a scatter of 0.26 which are over-plotted in Fig.~\ref{FIRC1400_bell}. Again, the distribution of our  $q^{\rm IR}~(1.4 \rm GHz)$  values  (with median = 2.61 and scatter = 0.30 for the main sample) has excellent agreement with that of Bell (2003). It is also worth noting that Bell (2003) found perfect agreement with the Yun et al. (2001) study, after correcting for the difference in the definitions of $q$ and $q^{\rm IR}$. Our results for the FIRC at 1.4 GHz are fully consistent with Yun et al. (2001)  and Bell (2003).  In the subsequent analysis, we will adopt the Bell (2003) definition of $q^{\rm IR}$ given in Eq. (10), based on the total IR to radio luminosity ratio. To calculate $q^{\rm IR}$ at 150 MHz, $q^{\rm IR} \ (150 \rm MHz)$, we can simply replace the 1.4-GHz luminosity $L_{\rm 1.4}$ with the 150-MHz luminosity $L_{150}$.

\subsection{The FIR-radio correlation at 150 MHz}

\begin{figure}
\centering
\includegraphics[height=2.6in,width=3.65in]{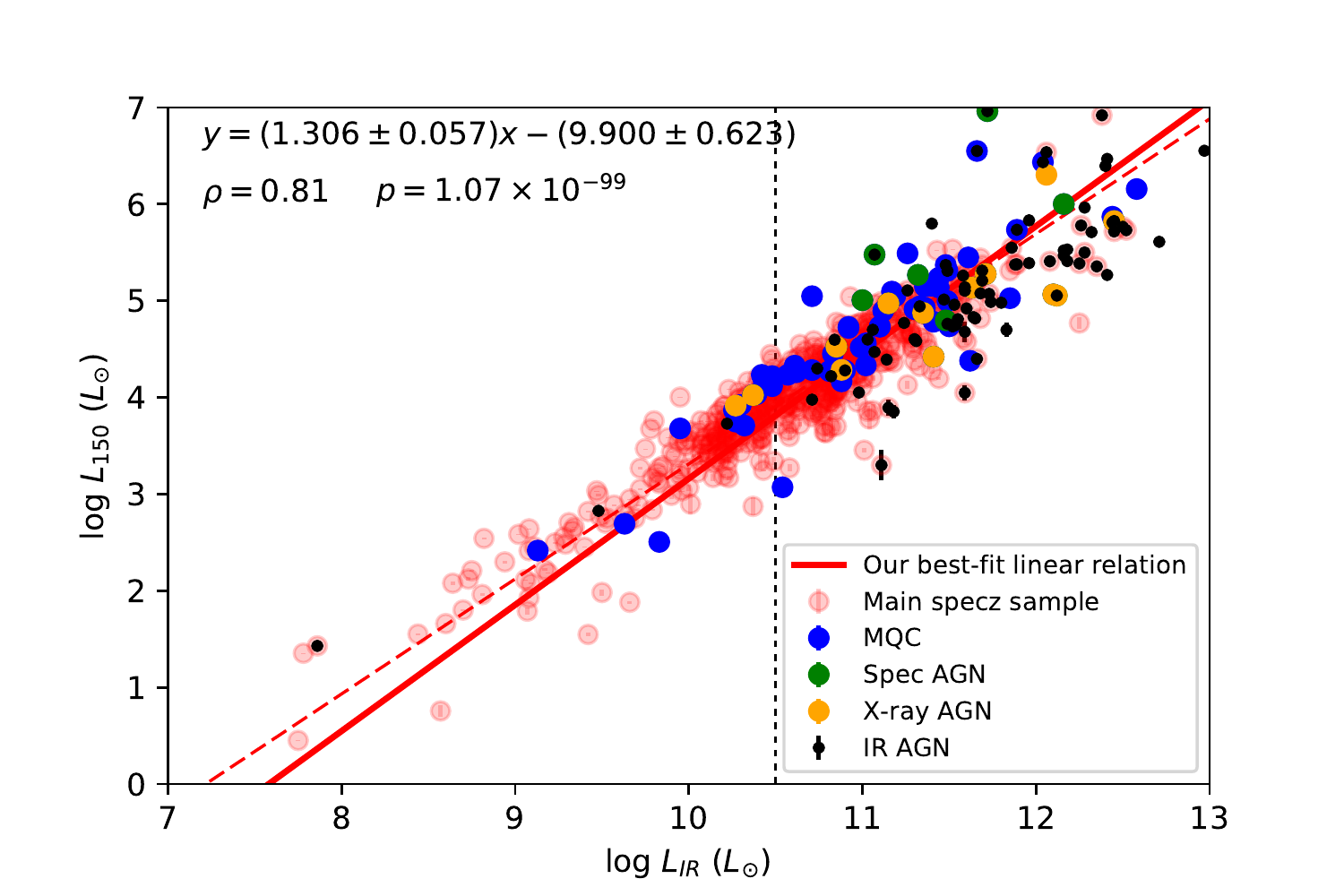}
\includegraphics[height=2.6in,width=3.65in]{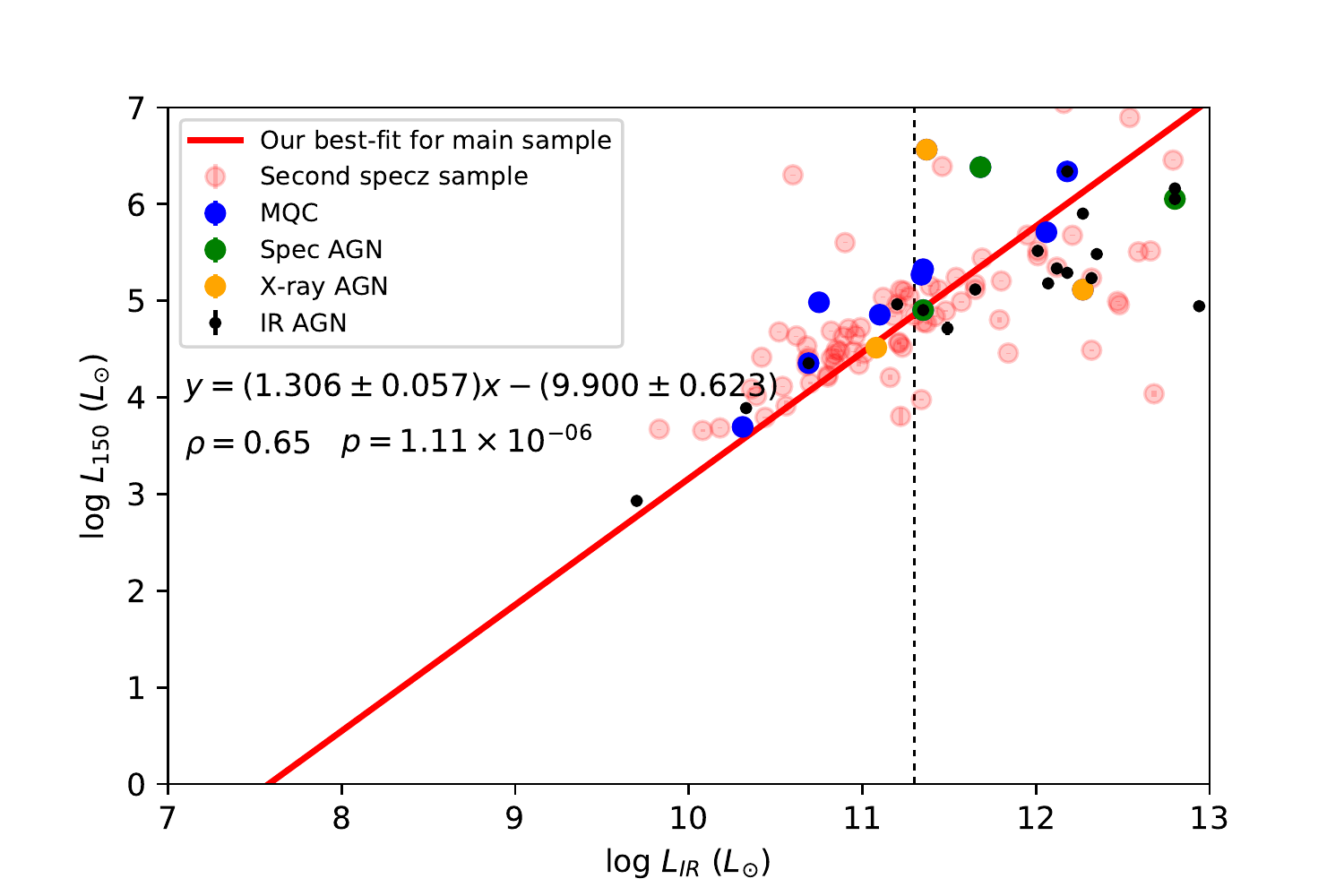}
\caption{Top: The correlation between the IR luminosity and the rest-frame 150-MHz luminosity for the main spec-$z$ sample, including AGNs identified in the X-ray, IR, the Million Quasar Catalog and in optical spectroscopy. The vertical dashed line indicates the 90\% completeness limit at the median redshift ($z\sim0.05$) of the main sample. Bottom: Same as the top panel but for the second sample. The vertical dashed line indicates the 90\% completeness limit at the median redshift ($z\sim0.12$) of the second sample.}
\label{LIR_vs_L150_agn}
\end{figure}

\begin{figure}
\centering
\includegraphics[height=2.6in,width=3.65in]{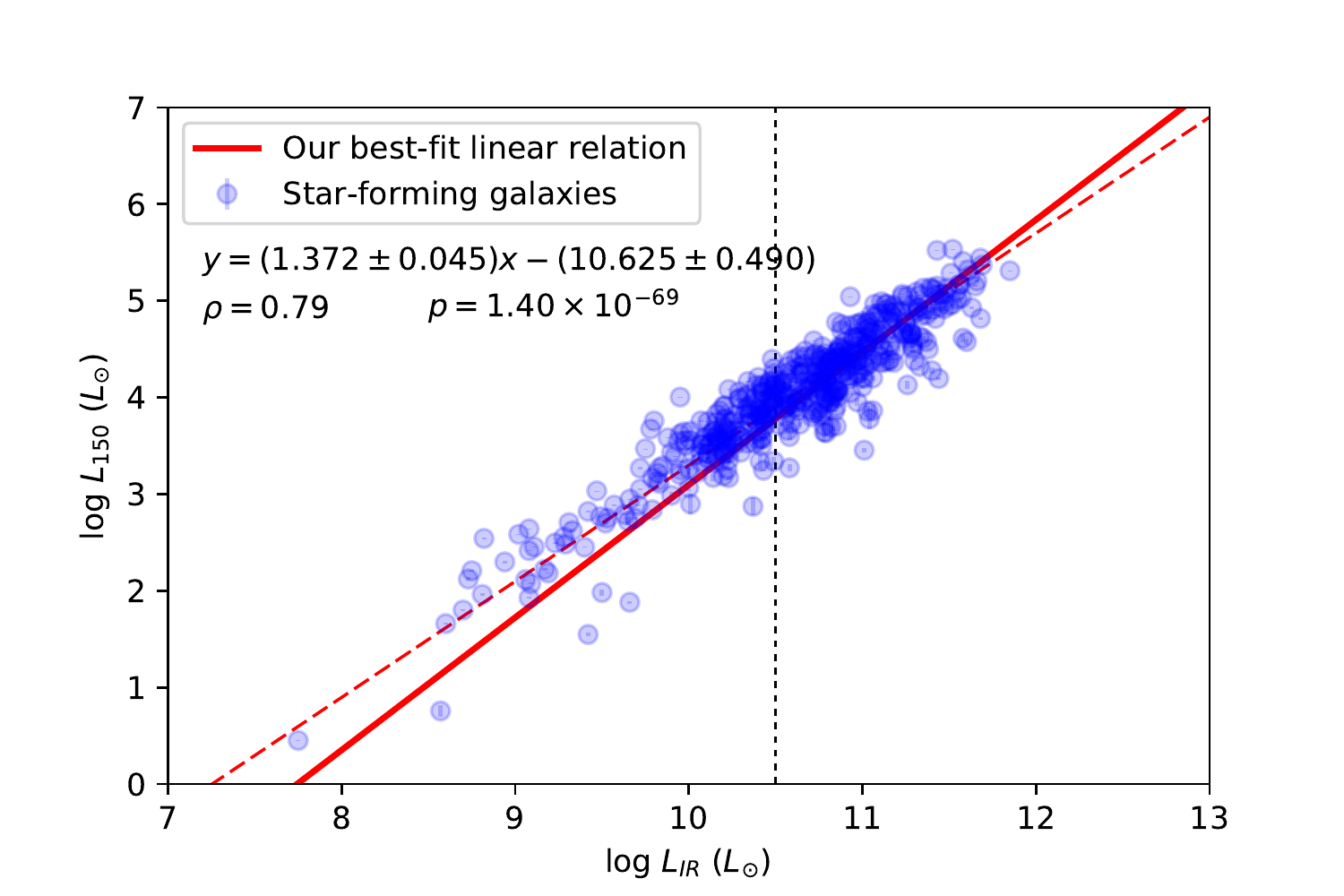}
\includegraphics[height=2.6in,width=3.65in]{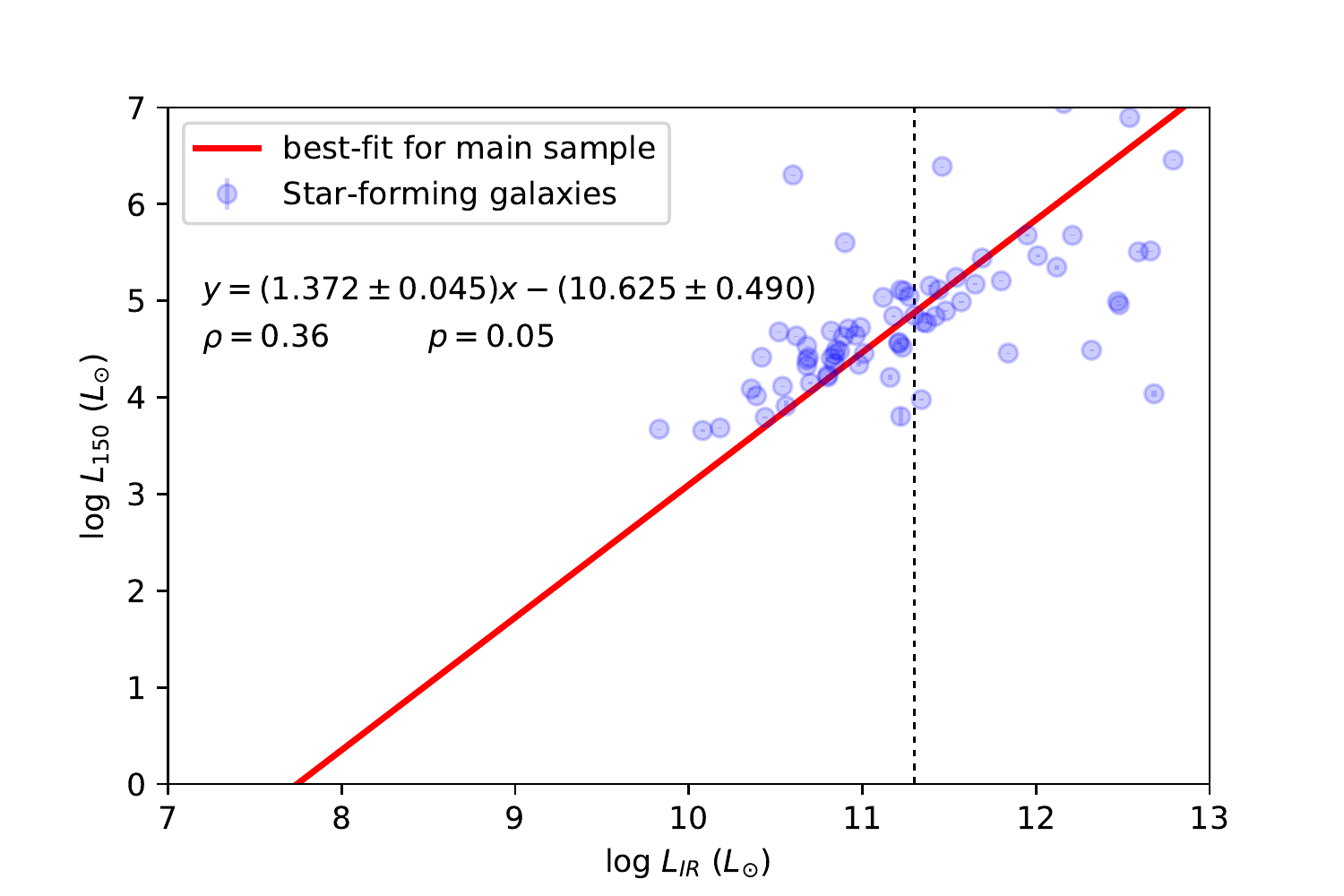}
\caption{Top: The correlation between the IR luminosity and the 150-MHz luminosity for the main sample, after excluding AGNs. The vertical dashed line indicates the 90\% completeness limit at the median redshift ($z\sim0.05$) of the main sample. Bottom: Same as the top panel but for the second sample. The vertical dashed line indicates the 90\% completeness limit at the median redshift ($z\sim0.12$) of the second sample.}
\label{LIR_vs_L150_sf}
\end{figure}

\begin{figure}
\centering
\includegraphics[height=2.6in,width=3.65in]{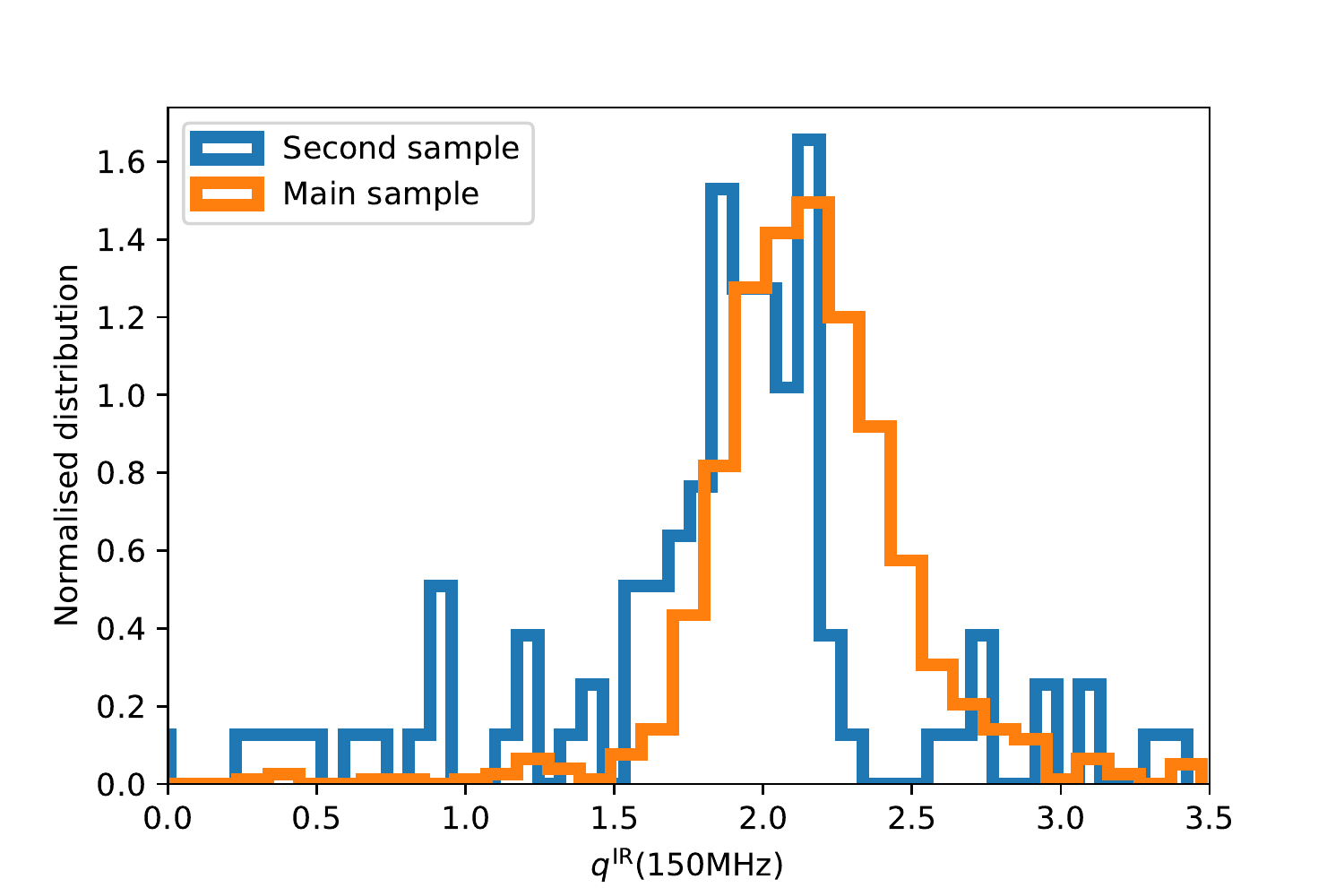}
\caption{Histogram of the $q^{\rm IR}$ values at 150 MHz using the definition in Eq. (10) and replacing the 1.4-GHz luminosity with the 150-MHz luminosity.}
\label{q150}
\end{figure}

\begin{figure}
\centering
\includegraphics[height=2.6in,width=3.65in]{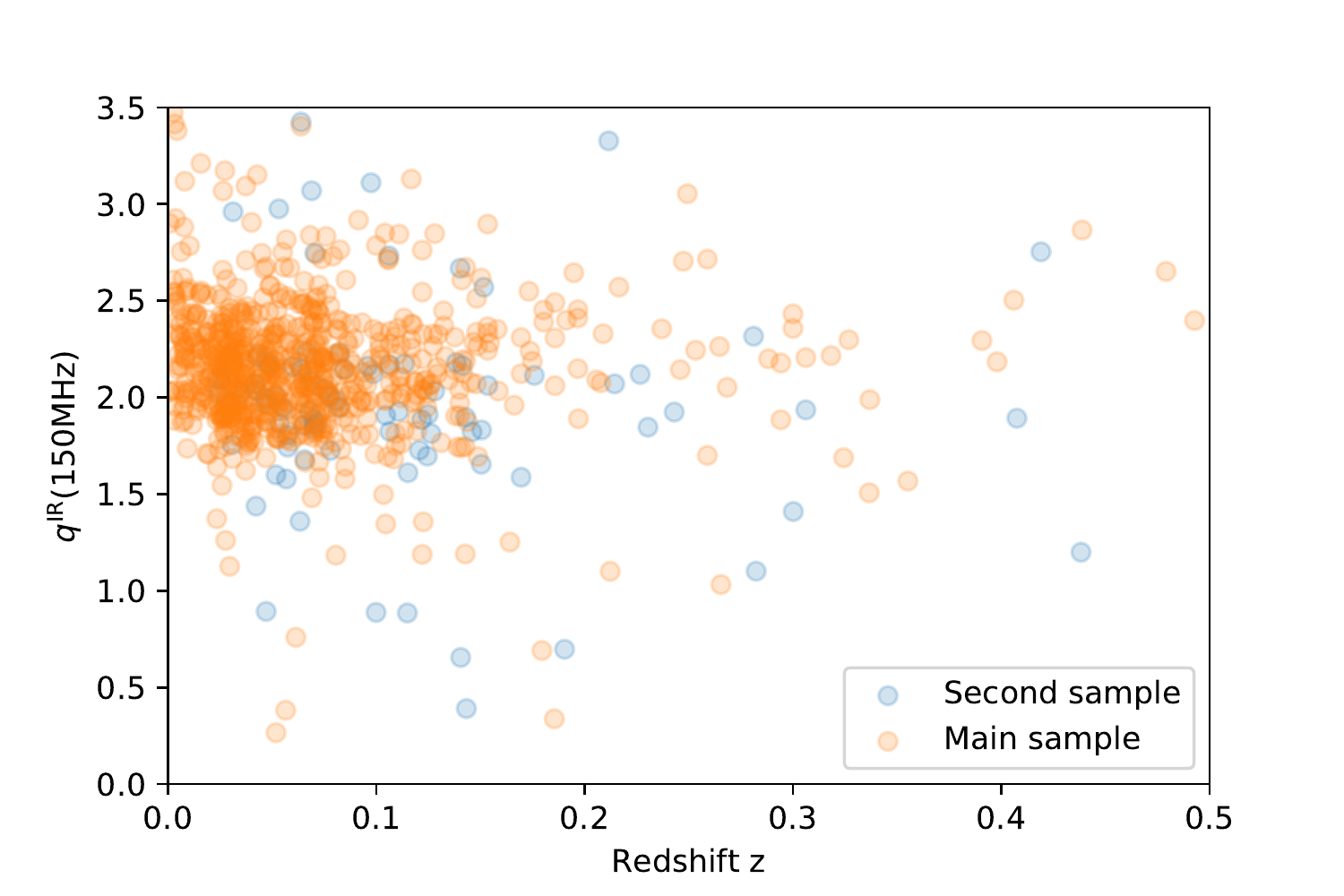}
\caption{The $q^{\rm IR} ~(150\rm MHz)$ values as a function of redshift for the RIFSCz-LOFAR matched sources.}
\label{qdep}
\end{figure}

\begin{table}
\centering
\caption{The numbers of AGNs identified by various methods in our main sample and second sample.}\label{table:agn_selection}
\begin{tabular}{ll}
\hline
\hline
Main sample&\\
\hline
AGN identification method & Number of sources \\
\hline
X-ray AGN & 13 \\
IR AGN & 84 \\
MQC AGN  & 71 \\
Spectroscopy AGN  & 16 \\
\hline
\hline
Second sample&\\
\hline
AGN identification method & Number of sources \\
\hline
X-ray AGN & 4 \\
IR AGN & 18 \\
MQC AGN  & 17 \\
Spectroscopy AGN  & 6 \\
\hline
\hline
\end{tabular}
\end{table}

Now we have shown that our results of the FIRC at 1.4 GHz are  consistent with previous measurements, we can study the FIRC at 150 MHz. First, to identify AGNs from our sample, we use the AGN classifications from the LOFAR VAC. As detailed in Duncan et al. (2018), AGN candidates have been identified using a variety of selection methods. Optical AGN are identified primarily through cross-matching with the Million Quasar Catalogue compilation of optical AGN, primarily based on SDSS (Adam et al. 2015) and other literature catalogues (Flesch 2015). Sources which have been spectroscopically classified as AGN are also flagged. Bright X-ray sources were identified based on the Second ROSAT all-sky survey (Boller et al. 2006) and the {\it XMM-Newton} slew survey.  Finally, IR AGNs are selected using the Assef et al. (2013) criteria based on magnitude and colour at the WISE W1 and W2 bands. We select sources with IRClass $>4$ from the VAC which corresponds to the ``75\% reliability'' selection criteria.  Table 3 lists the number of identified AGNs in our samples.

The top panel in Fig.~\ref{LIR_vs_L150_agn} shows the correlation between $\log L_{\rm IR}$ and the rest-frame 150MHz luminosity  $\log L_{150}$ for the main spec-$z$ sample and AGNs (predominantly luminous systems) identified using X-ray, optical and IR data. The vertical dashed line indicates the 90\% completeness limit at the median redshift ($z\sim0.05$) of the main sample, at $L_{\rm IR}\sim10^{10.5}L_{\odot}$. This value is derived from multiplying the 90\% completeness limit at 60 $\mu$m, $L_{60}\sim10^{10.27}L_{\odot}$, by the median ratio of $L_{\rm IR}$ to $L_{60}$ using the IR SED templates from Chary \& Elbaz (2001). The Chary \& Elbaz (2001) templates are shown to be able to reproduce the observed luminosity-luminosity correlations at various IR wavelengths for local galaxies. In comparison, the selection effect due to the median sensitivity (71 $\mu$Jy/beam) of the LOFAR 150-MHz observations is negligible (i.e., LOFAR is much deeper than IRAS for typical galaxy SEDs). At $z\sim0.05$, this median sensitivity corresponds to $ L_{150} = 10^{2.92} L_{\odot}$ at $5\sigma$. We perform a linear regression which is based on a fitting method called the bivariate correlated errors and intrinsic scatter (BCES) described in Akritas \& Bershady (1996). We use the public code developed in Nemmen et al. (2012).  The red solid line shows our best-fit linear relation using galaxies above the 90\% completeness limit, 
\begin{equation}
\log L_{150} \ (L_{\odot}) = 1.306~(\pm0.057) \times \log L_{\rm IR} \ (L_{\odot}) - 9.900~(\pm0.623),
\end{equation}
while the red dashed line shows the best-fit relation using all galaxies. While some optically-identified AGNs clearly show an excess radio emission and therefore do not lie on the FIRC, most of the optical AGNs still obey the FIRC. Most of the IR and X-ray identified AGN also lie on the FIRC. 

The bottom panel in Fig.~\ref{LIR_vs_L150_agn} shows the correlation between $\log L_{\rm IR}$ and $\log L_{150}$ for the second sample. The vertical dashed line indicates the 90\% completeness limit at the median redshift ($z\sim0.12$) of the second sample, at $L_{\rm IR}\sim10^{11.3}L_{\odot}$. By comparison, the LOFAR sensitivity limit at $z\sim0.12$ is at around $L_{150} = 10^{3.71} L_{\odot}$ at $5\sigma$. We do not attempt to fit the second sample (due to the small sample size) but simply over-plot the best-fit linear relation for the main sample which seems to describe the second sample reasonably well.

The top panel in Fig.~\ref{LIR_vs_L150_sf}  shows the correlation between $\log L_{150}$ and $\log L_{\rm IR}$ for our star-forming galaxies from the main sample, after removing AGNs. Using the BCES method, our best-fit linear relation between the log of $L_{150}$  and the log of $L_{\rm IR}$ for galaxies above the 90\% completeness limit (plotted as the red solid line) is,
\begin{equation}
\log L_{150}  \ (L_{\odot}) = 1.372~(\pm0.045) \times \log L_{\rm IR}  \ (L_{\odot}) - 10.625~(\pm0.490).
\end{equation}
The best-fit relation derived for all galaxies is plotted as the red dashed line. We also test the significance of the correlation by calculating the Pearson correlation coefficient $\rho$ which is found to be 0.79 and the p-value which is $1.40\times10^{-69}$. The bottom panel in Fig.~\ref{LIR_vs_L150_sf}  shows the correlation between $\log L_{\rm IR}$ and $\log L_{150}$ for star-forming galaxies in the second sample. Again we do not fit the second sample but simply over-plot the best-fit linear relation for the main sample. The Pearson correlation coefficient $\rho$ and p-value for galaxies above the 90\% completeness limit in the second sample are 0.36 and $0.05$, respectively.

Fig.~\ref{q150} shows the distribution of $q^{\rm IR}$ (150 MHz) values of our sample derived using Eq. (10) and replacing the 1.4-GHz luminosity with the 150-MHz luminosity. The median value and scatter of $q^{\rm IR}$ (150 MHz) are 2.14 and 0.34, respectively, for the main sample. The median value and scatter are 1.93 and 0.61, respectively, for the second sample. Calistro-Rivera et al. (2017) found a median $q^{\rm IR}$ (150 MHz) value of 1.544. This is inconsistent with our result. The main cause of this inconsistency is the large difference in the distributions of $L_{\rm IR}$ in the two studies. The mean $L_{\rm IR}$ of the galaxy sample in Calistro-Rivera et al. (2017) is roughly 1.3 dex higher than this study. Using Eq. (12), we can derive that an increase in $L_{\rm IR}$ by 1.3 dex would reduce  $q^{\rm IR}$ (150 MHz)  by $\sim0.5$.

In Fig.~\ref{qdep}, we plot the $q^{\rm IR} ~(\rm 150MHz)$ values against redshift. A mild redshift evolution has been report by Calistro Rivera et al. (2017) and Read et al. (2018). We do not see significant evidence for any redshift evolution although our sample is perhaps too low redshift to see any evolutionary effects.  When LoTSS is completed, the areal overlap between IRAS and LoTSS will reach $\sim20,000$ deg$^2$. By then, we will have a much larger cross-matched sample which will be more adequate for detecting mild redshift evolution effect, if it exists.

\subsection{The rest-frame 150-MHz luminosity as a SFR tracer}

\begin{figure*}
\centering
\includegraphics[height=3.5in,width=4.9in]{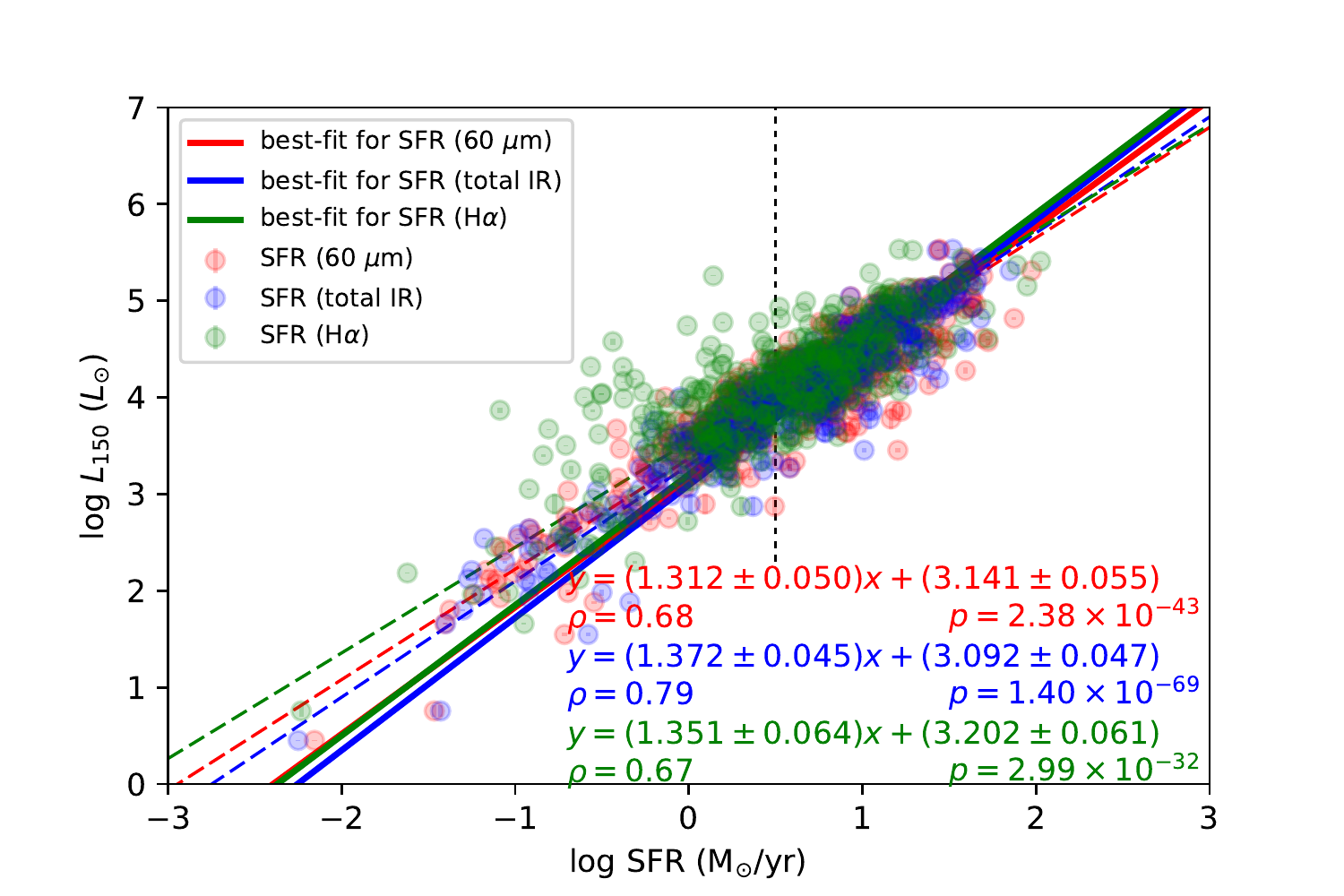}
\includegraphics[height=3.5in,width=4.9in]{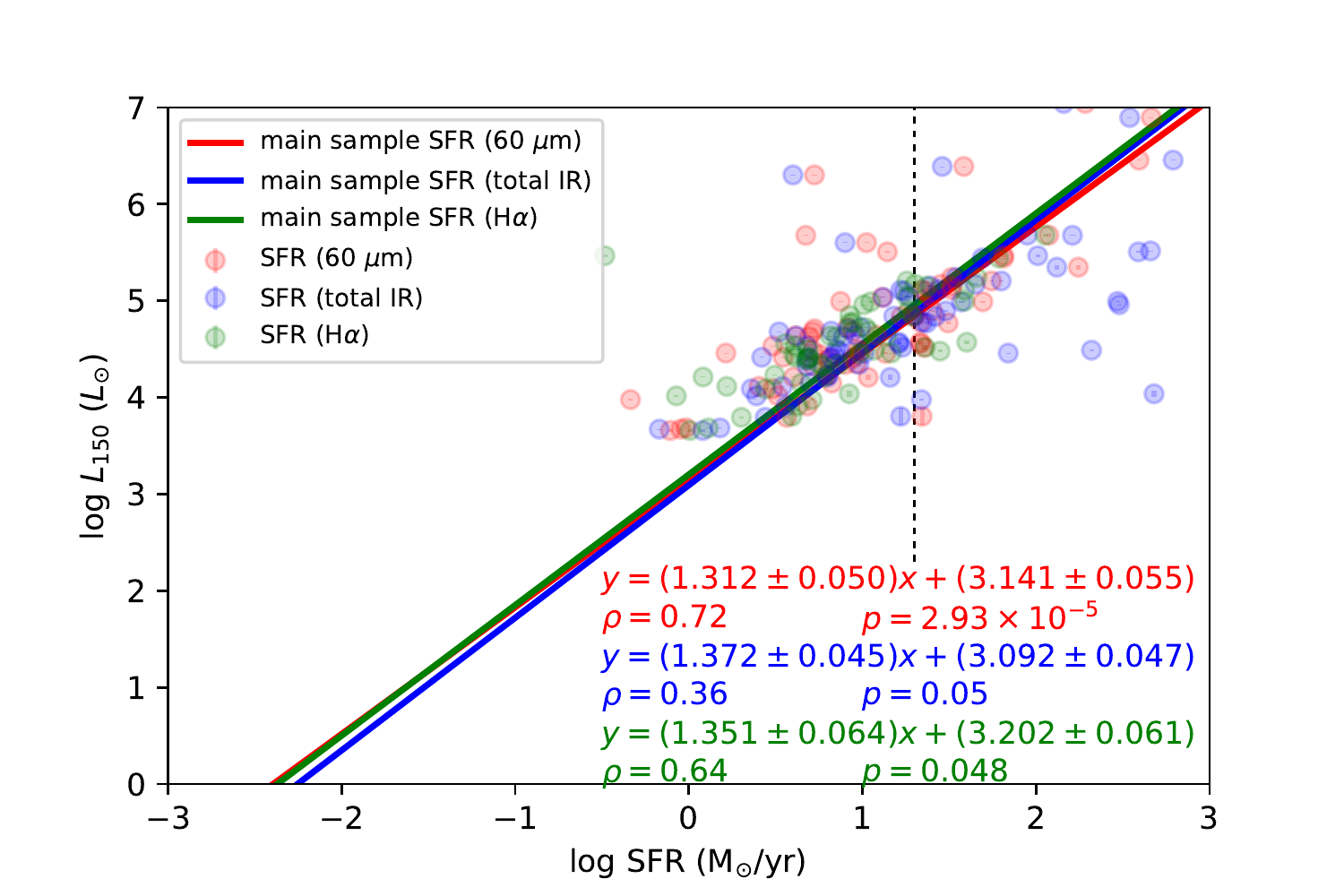}
\caption{Top: The correlation between the rest-frame 150-MHz luminosity and various SFR tracers for the main sample, after excluding AGNs. The vertical dashed line indicates the 90\% completeness limit at the median redshift ($z\sim0.05$) of the main sample. The solid lines are best-fit relations derived using only galaxies above the completeness limit. The dashed lines are best-fit relations derived using all galaxies. Bottom: Same as the top panel but for the second sample. The vertical dashed line indicates the 90\% completeness limit at the median redshift ($z\sim0.12$) of the second sample.}
\label{SFR_vs_L150}
\end{figure*}

In the top panel in Fig.~\ref{SFR_vs_L150}, we compare the rest-frame 150-MHz luminosity $L_{150}$ with several SFR tracers for star-forming galaxies from the main sample. The blue symbols correspond to SFRs derived based on the total IR luminosity $L_{\rm IR}$. The red symbols correspond to SFRs provided in the RIFSCz based on $L_{60}$ (see Section 2.1). The green symbols correspond to SFR derived from the H$\alpha$ line luminosity. Good agreement between the various SFR estimates are found. Our best-fit linear relation between $\log L_{150}$ and the logarithmic value of SFR based on $L_{60}$ for galaxies above the 90\% completeness limit is,
\begin{equation}
\begin{aligned}
\log L_{150} \ (L_{\odot}) = {} & 1.312~(\pm0.050)\times \log {\rm SFR_{60}} \ (M_{\odot} /{\rm yr}) \\
&+ 3.141 ~(\pm 0.055).
\end{aligned}
\end{equation}
The Pearson correlation coefficient $\rho$ is equal to 0.68 and the p-value is $2.38\times10^{-43}$. Our best-fit linear relation between $\log L_{150}$ and the logarithm of SFR based on $L_{\rm IR}$ for galaxies above the 90\% completeness limit is,
\begin{equation}
\begin{aligned}
\log L_{150}  \ (L_{\odot}) = {} & 1.372~(\pm0.045) \times \log {\rm SFR_{IR}} \ (M_{\odot} /{\rm yr}) \\
&+ 3.092 ~(\pm 0.047).
\end{aligned}
\end{equation} 
The Pearson correlation coefficient $\rho$ is equal to 0.79 and the p-value is $1.40\times10^{-69}$.  Our best-fit linear relation between $\log L_{150}$ and the logarithm of SFR based on H$\alpha$ line luminosity for galaxies above the 90\% completeness limit is,
\begin{equation}
\begin{aligned}
\log L_{150}  \ (L_{\odot}) = {} & 1.351~(\pm0.064) \times \log {\rm SFR_{H\alpha}} \ (M_{\odot} /{\rm yr}) \\
&+ 3.202~ (\pm 0.061).
\end{aligned}
\end{equation} 
The Pearson correlation coefficient $\rho$ is equal to 0.67 and the p-value is $2.99\times10^{-32}$. Thus, the relation between the logarithm of the 150-MHz luminosity and the logarithm of SFR is linear with a slope of 1.3 over a dynamic range of four orders of magnitude in SFR.  We also show the best-fit relations derived using all galaxies, i.e., including the fainter galaxies below the completeness limit. These relations (plotted as dashed lines) show shallower slopes.

The bottom panel in Fig.~\ref{SFR_vs_L150} compares $L_{150}$ with several SFR tracers for star-forming galaxies from the second sample. Due to the small sample size, we do not attempt to fit the second sample but simply over-plot the best-fit linear relations for the main sample. In the plot, we also show the Pearson correlation coefficient $\rho$ and p-value derived for the galaxies above the 90\% completeness limit in the second sample.

\section{{\bf Conclusions}}

In this paper, we set out to study the FIRC in both the 1.4-GHz and the 150-MHz bands in the local Universe as the median redshift of our main sample is at $z\sim0.05$, with the aim of testing the use of the rest-frame 150-MHz luminosity $L_{150}$ as a SFR tracer. We cross-match the 60-$\mu$m selected RIFSCz catalogue and the 150-MHz selected LOFAR VAC in the HETDEX spring field, using a combination of the closest match method and the likelihood ratio technique. We also cross-match our sample with the 1.4-GHz selected FIRST survey catalogue. We estimate $L_{150}$ for the LOFAR sources and compare it with the IR luminosity, $L_{\rm IR}$, and several SFR tracers, after removing AGNs. Our main conclusions are:

\begin{itemize}

\item A linear and tight correlation with a slope of unity between $\log L_{\rm IR}$ and $\log L_{\rm 1.4}$ holds. Our median $q$ value  and scatter at 1.4 GHz for the main sample, which are 2.37 and 0.26, respectively, are consistent with previous studies such as Yun et al. (2001).

\item A linear and tight correlation between $\log L_{\rm IR}$ and $\log L_{150}$  holds with a slope of 1.37. Our median $q^{\rm IR}$ value is higher than the number reported in Calistro Rivera et al. (2017). This is mainly due to a large difference in the distributions of $L_{\rm IR}$ of our samples.

\item The logarithm of $L_{150}$ correlates tightly with the logarithm of SFR derived from three tracers, including SFR derived from H$\alpha$ line luminosity, the rest-frame 60-$\mu$m luminosity and $L_{\rm IR}$.  Best-fit formulae for the correlation between $L_{150}$ and the three SFR tracers are provided, which are in excellent agreement with each other. The logarithmic slope ($\sim1.3$) of the correlation between $L_{150}$ and SFR suggests that the correlation is non-linear.

\end{itemize}

The LoTSS Second Data Release will include images and catalogues for 2,500 deg$^2$ of the northern sky and will be released by 2020. The all-sky IRAS survey allows the maximum areal overlap with LOFAR. At the eventual completion of LoTSS, the areal overlap between IRAS and LoTSS will reach $\sim$ 20,000 deg$^2$. Therefore, we will be able to not only repeat the same analysis with a much larger sample but also to study in detail the FIRC at 150 MHz and its variation with galaxy physical properties such as stellar mass, SED type and morphology.

\begin{acknowledgements}
We thank the anonymous referee for a through and constructive report. We thank Rainer Beck for helpful discussions on the far-infrared to radio correlation. SCR acknowledges support from the UK Science and Technology Facilities Council [ST/N504105/1]. MB and IP acknowledge support from INAF under PRIN SKA/CTA FORECaST. MJH acknowledges support from the UK Science and Technology Facilities Council [ST/R000905/1]. JS is grateful for support from the UK Science and Technology Facilities Council (STFC) via grant ST/M001229/1 and ST/R000972/1. GG acknowledges the postdoctoral research fellowship from CSIRO. HJAR, WLW and KJD acknowledge support from the ERC Advanced Investigator programme NewClusters 321271. WLW also acknowledges support from the CAS-NWO programme for radio astronomy with project number 629.001.024, which is financed by the Netherlands Organisation for Scientific Research (NWO).

Funding for SDSS-III has been provided by the Alfred P. Sloan Foundation, the Participating Institutions, the National Science Foundation, and the U.S. Department of Energy Office of Science. The SDSS-III web site is http://www.sdss3.org/.
SDSS-III is managed by the Astrophysical Research Consortium for the Participating Institutions of the SDSS-III Collaboration including the University of Arizona, the Brazilian Participation Group, Brookhaven National Laboratory, Carnegie Mellon University, University of Florida, the French Participation Group, the German Participation Group, Harvard University, the Instituto de Astrofisica de Canarias, the Michigan State/Notre Dame/JINA Participation Group, Johns Hopkins University, Lawrence Berkeley National Laboratory, Max Planck Institute for Astrophysics, Max Planck Institute for Extraterrestrial Physics, New Mexico State University, New York University, Ohio State University, Pennsylvania State University, University of Portsmouth, Princeton University, the Spanish Participation Group, University of Tokyo, Uni- versity of Utah, Vanderbilt University, University of Virginia, University of Washington, and Yale University.

LOFAR is the Low Frequency Array designed and constructed by ASTRON. It has observing, data processing, and data storage facilities in several countries, which are owned by various parties (each with their own funding sources), and which are collectively operated by the ILT foundation under a joint scientific policy. The ILT resources have benefitted from the following recent major funding sources: CNRS-INSU, Observatoire de Paris and Universit\'{e} d'Orl\'{e}ans, France; BMBF, MIWF-NRW, MPG, Germany; Science Foundation Ireland (SFI), Department of Business, Enterprise and Innovation (DBEI), Ireland; NWO, The Netherlands; The Science and Technology Facilities Council, UK; Ministry of Science and Higher Education, Poland.

This research made use of the Dutch national e-infrastructure with support of the SURF Cooperative (e-infra 180169) and the LOFAR e-infra group. The J\"{u}lich LOFAR Long Term Archive and the German LOFAR network are both coordinated and operated by the J\"{u}lich Supercomputing Centre (JSC), and computing resources on the Supercomputer JUWELS at JSC were provided by the Gauss Centre for Supercomputing e.V. (grant CHTB00) through the John von Neumann Institute for Computing (NIC).

This research made use of the University of Hertfordshire high-performance computing facility and the LOFAR-UK computing facility located at the University of Hertfordshire and supported by STFC [ST/P000096/1].

\end{acknowledgements}

\end{document}